# Segment-Sliding Reconstruction of Pulsed Radar Echoes with Sub-Nyquist Sampling

Suling Zhang, Feng Xi, *Member, IEEE*, Shengyao Chen, *Member, IEEE*,
Yimin D. Zhang, *Senior Member, IEEE*, and Zhong Liu, *Member, IEEE*

***Abstract*—It has been shown that analog-to-information conversion (AIC) is an efficient scheme to perform sub-Nyquist sampling of pulsed radar echoes. However, it is often impractical, if not infeasible, to reconstruct full-range Nyquist samples because of huge storage and computational load requirements. Based on the analyses of AIC measurement system, this paper develops a novel segment-sliding reconstruction (SegSR) scheme to effectively reconstruct the Nyquist samples. The SegSR performs segment-by-segment reconstruction in a sliding mode and can be implemented in real-time. An important characteristic that distinguish the proposed SegSR from the existing methods is that the measurement matrix in each segment satisfies the restricted isometry property (RIP) condition. Partial support in the previous segment can be incorporated into the estimation of the Nyquist samples in the current segment. The effect of interference introduced from adjacent segments is theoretically analyzed, and it is revealed that the interference consists of two interference levels having different impacts to the signal reconstruction performance. With these observations, a two-step orthogonal matching pursuit (OMP) procedure is proposed for segment reconstruction, which takes into account different interference levels and partially known support of the previous segment. The proposed SegSR achieves nearly optimal reconstruction performance with a significant reduction of computational loads and storage requirements. Theoretical analyses and simulations verify its effectiveness.**

## I. INTRODUCTION

A pulsed radar usually transmits modulated pulses during the transmit time and receives during the receive time any echo signals reflected from illuminated targets. The received echoes are sampled and processed to extract target information. To avoid loss of information, the Nyquist sampling theorem requires the sampling rate to be higher than twice the radar signal bandwidth. Many applications require the radar systems to employ wideband signals in order to achieve a fine range resolution. In such a situation, the large bandwidth mandates the use of high-rate analog-to-digital converter (ADC) and thus yields a large volume of sampled data which become highly demanding in the subsequent stages in terms of both storage and processing capabilities. The requirement for the ADC devices to operate in the Nyquist sampling rate has become a serious bottleneck in the development of wideband and ultra-wideband radar systems.

The recent advances of compressive sampling (CS) [1]-[4], or compressive sensing, open a new avenue to achieve data acquisition with a significantly reduced sampling rate. The CS theory exploits the sparsity of signals and samples them in a rate that is close to their information rate rather than their bandwidth. With a high probability, CS techniques can recover sparse signals from far fewer samples or measurements than those that are determined by the Nyquist sampling theorem. In this case, the required sampling rate is reduced, leading to a simplified ADC configuration. A number of schemes have been proposed to implement the CS or analog-to-information conversion (AIC) of analog signals, such as random demodulation (RD) [5], [6], segmented compressed sampling [7], random modulator pre-integrator (RMPI) [8], Xampling [9], and quadrature compressed sensing (QuadCS) [10], [11]. Theoretical analyses and experimental studies have shown that these AIC systems are efficient for low-rate (sub-Nyquist) acquisition of sparse signals with a large bandwidth.

The AIC system have attracted wide attention for radar signal acquisition and processing [12]-[16]. Reference [12] reports an RMPI-based radar pulse receiver in which the target information is extracted without reconstructing the full Nyquist samples. The QuadCS-based pulse-Doppler processing in [13] reconstructs target information from in-phase and quadrature sub-Nyquist samples. A Doppler-focusing approach based on the Xampling system is developed in [14]. Reference [15] proposes an adaptive compressive sensing and processing method for radar tracking. More references can be found, e.g., in a recent workshop on CS radar [16].

In many cases, we desire to recover the Nyquist samples of radar echoes from the low-rate outputs of AIC systems. In CS theory, the problem refers to sparse signal recovery or reconstruction, which is often mathematically expressed as

$$\begin{cases} \min_{\boldsymbol{\sigma}} \|\boldsymbol{\sigma}\|_0 \\ \text{s.t. } \mathbf{y} = \boldsymbol{\Phi}\boldsymbol{\Psi}\boldsymbol{\sigma} = \mathbf{A}\boldsymbol{\sigma}, \end{cases} \tag{1}$$

where $\|\boldsymbol{\sigma}\|_0$ denotes the $\ell_0$-norm of $\boldsymbol{\sigma}$ which counts for the number of nonzero entries, i.e., the sparsity, of the $N\times 1$ sparse coefficient vector $\boldsymbol{\sigma}$, $\mathbf{y}$ is an $M\times 1$ measurement vector, $\boldsymbol{\Phi}$ is an $M\times N$ observation matrix with $M \ll N$, $\boldsymbol{\Psi}$ is an $N\times N$

Manuscript received XXXXXX. This work was supported in part by the National Science Foundation of China under Grant 61171166 and 61101193.

S. Zhang, F. Xi, S. Chen and Z. Liu are with the Department of Electrical Engineering, Nanjing University of Science and Technology, Nanjing, 210094, China (e-mail: zsljust@163.com, xifeng.njust@gmail.com, chen_shengyao@163.com, and eezliu@njust.edu.cn).

Y. D. Zhang is with the Center for Advanced Communications, Villanova University, Villanova, PA 19085 USA (e-mail: yimin.zhang@villanova.edu).

basis matrix, and $\mathbf{A}=\mathbf{\Phi\Psi}$ is the yielding $M\times N$ measurement matrix. The sparsity of $\boldsymbol{\sigma}$ is denoted as $\|\boldsymbol{\sigma}\|_0 = K$. It has been shown that the sparse vector $\boldsymbol{\sigma}$ can be exactly reconstructed by solving (1), if matrix $\mathbf{A}$ satisfies the restricted isometry property (RIP) condition [3], [4],

$$(1-\delta_K)\|\boldsymbol{\sigma}\|_2^2 \le \|\mathbf{A}\boldsymbol{\sigma}\|_2^2 \le (1+\delta_K)\|\boldsymbol{\sigma}\|_2^2, \tag{2}$$

where $\delta_K$ is a sufficiently small RIP constant. Unfortunately, problem (1) is a combinatorial NP-complete problem and computationally intractable [17]. In practice, therefore, the $\ell_0$-norm used in (1) is often replaced by the $\ell_1$-norm, yielding the following convex optimization problem,

$$\begin{cases} \min\limits_{\boldsymbol{\sigma}} \|\boldsymbol{\sigma}\|_1 \\ \text{s.t. } \mathbf{y}=\mathbf{\Phi\Psi}\boldsymbol{\sigma}=\mathbf{A}\boldsymbol{\sigma}. \end{cases} \tag{3}$$

There are a wide variety of approaches to solve (3), including the greedy iteration algorithms [18], [19], convex optimization algorithms [20], [21] and Bayesian CS [21], [23] (see [24] for a review). All these techniques are suitable when the measurement matrix is of a moderate size. For large-scale measurement, on the other hand, we need to allocate a prohibitively huge size of memory space to store the data, as such, the recovery may become impractical for the state-of-the-art computers. Consider, for example, a radar system with a signal bandwidth of $100MHz$, a pulse width of $10\mu s$, and a receiving time of $2490\mu s$. At one-fifth of the Nyquist sampling rate, we need to store about $1.24\times10^{10}$ elements with 49800 rows and 248000 columns, occupying about 92GB of memory using the standard IEEE double precision. As a result, the full-range reconstruction is impractical, if not infeasible, with the state-of-the-art hardware capabilities.

Motivated by such facts, this paper studies implementable, full range reconstruction of pulsed radar echoes sampled by an RD AIC system. The results can be easily extended to other sparse sampling schemes likes RMPI, QuadCS and Xampling that generate similar measurement matrices.

For the RD AIC system, as will be discussed in Section III, its measurement matrix $\mathbf{A}$ with the waveform-matched dictionary [25] is sparse with its effective elements around the diagonal and there are overlaps between the rows, as shown in Fig. 1. With this specific matrix structure, we can decompose the large-scale reconstruction problem (3) into a series of small-scale ones by properly segmenting the measurement vector and the sparse vector. The word "*properly*" implies that the segmentations are made such that each segmented measurement matrix satisfies the RIP conditions. In this case, the entire sparse vector $\boldsymbol{\sigma}$ can be reconstructed segment by segment.

It is important to note that, because matrix $\mathbf{A}$ is not block-diagonal, the reconstruction of a segment will be interfered by the adjacent segments. The interference is referred to as *virtual noise* in this paper and is factorized into forward virtual noise and backward virtual noise. The forward virtual noise is generated by the inaccurate estimate of the sparse vector in the previous segment, whereas the backward one is formulated by partial measurement in the subsequent segment. We theoretically analyze the effects of the virtual noise on reconstructed positions and amplitudes of the sparse entries in current segment. It is revealed that the virtual noise has different levels of distributions for different measurements, and the effect of the backward virtual noise is much higher than that of the forward one.

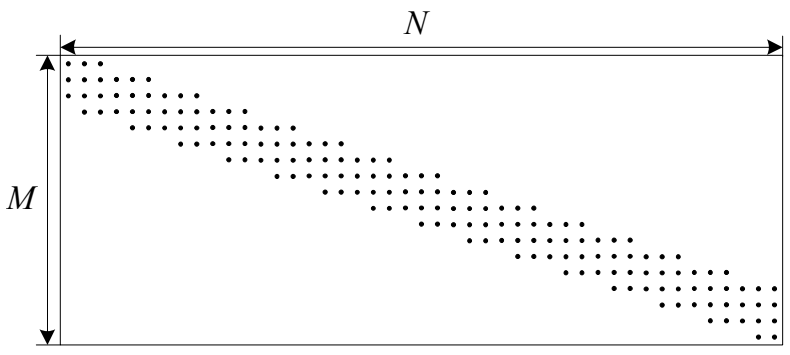

Fig. 1. The structure of measurement matrix.

To perform efficient estimation of sparse echoes in each segment, we develop a two-step OMP process (TOMPP) algorithm which takes into account the effects of these two kinds of virtual noises and partially known support in the previous segment estimate. In the proposed approach, we take the segment-sliding way to segment both the measurement vector and the sparse vector. As a result, the overall complexity is reduced, and the full-range reconstruction of the pulsed radar echoes can be implemented in real time. Theoretical analyses and simulation results validate the effectiveness of the proposed segment estimation. When the background noise is Gaussian, the proposed scheme achieves nearly optimal reconstruction performance with significant reduction of computational loads and storage requirements.

The problem studied in this paper is in essence a reconstruction of sparse time-varying signals from streaming measurements, which have been discussed in [26]-[30]. These algorithms fully exploit the measurement systems and perform sliding estimation of the sparse coefficients. The methods in [26], [27] use a sliding window for data processing and entail recursive sampling and iterative recovery. In each sliding sampling, the measurement matrix is recursively generated by permuting the previous measurement matrix. The measurement matrix in each sampling always satisfies the RIP condition if the original one satisfies. The method in [28] focuses on the measurement of infinite-dimensional signals using a finite-length, time-varying linear system. The compressive measurement over a finite-length window is closely linked with a segment of the infinite-dimensional signal. Then the signal recovery can be implemented iteratively over sliding intervals. Similar idea is taken in [29] with special lapped orthogonal transform bases. Reference [30] studies the large-scale reconstruction problem arising from radar imaging, and an interesting segmentation approach is presented.

There are two important issues that are not properly addressed in [28]-[30]. First, because matrix $\mathbf{A}$ is not block-diagonal, the reconstruction of a segment will be interfered by the adjacent segments. The effect of such interference, however, has not been properly studied and analyzed. Second, the existing approaches do not guarantee the RIP conditions to be satisfied after the segmentation, even when the original problem satisfies the RIP condition. With different segmentation and thorough theoretical analyses, our works well solve above problems. Our main contributions include: (a) We develop a new segmentation-based algorithm in which all the measurement sub-matrices satisfy the RIP condition; (b) We provide a theoretical analysis of the effect of the interferences and reveal the interference structure consisting of two interference noise levels; and (c) Based on the above results, a new OMP algorithm, TOMPP, is exploited for segment reconstruction, which takes into account the impacts of different noise levels to the signal reconstruction performance. The TOMPP algorithm achieves almost optimized reconstruction performance.

The remainder of the paper is organized as follows. In Section II, we provide a brief statement of problem model. In Section III, we analyze the structure of the measurement matrix and explain the mechanism of sliding reconstruction. Section IV describes the proposed segment-sliding reconstruction algorithm. An analysis of the effect of the interference from adjacent segments is provided in Section V. The required storage capacity and computational complexity are analyzed in Section VI. Numerical results are presented in Section VII. We conclude this paper in Section VIII. The proofs of the main results are given in the Appendix.

*Notations*: Bold letters denote the vectors or matrices. $(\cdot)^T$ and $(\cdot)^H$ represent transposition and conjugate transposition of a vector or matrix, respectively. $(\cdot)^\dagger$ represents the Moore-Penrose inverse of a tall, full-rank matrix. $\|\cdot\|_1$, $\|\cdot\|_2$ and $\|\cdot\|_\infty$ denote the $\ell_1$, $\ell_2$ and $\ell_\infty$ vector or matrix norm, respectively. $|\cdot|$ represents absolute value of a number or cardinality of a set. A vector (matrix) with a set as its subscript denotes the sub-vector (sub-matrix) containing the elements (columns) of the vector (matrix) indexed by the set.

## II. Background Materials and Problem Statement

### A. Signal Model

Consider a pulsed radar where the baseband signal $s(t)$ has a pulse width of $T_p$ and a band width of $B/2$. Then, for $K$ non-fluctuating point targets, the received echo signal at the base-band can be represented as

$$x(t)=\sum_{k=0}^{K-1}\sigma_k s(t-t_k),\quad t\in[0,T),\tag{4}$$

where $t_k$ and $\sigma_k$ are the time delay and gain coefficient of the $k$-th target, respectively, and $T$ refers to the receive time which is usually much larger than $T_p$, i.e., $T\gg T_p$. The Nyquist sampling of $x(t)$ during the receive interval $T$ yields at least $BT$ samples. For notational simplicity, the background noise is not included in the above model, but its effect on the reconstruction performance will be examined numerically in Section VII.

To obtain the sub-Nyquist samples of $x(t)$ to be processed by the CS theory, $x(t)$ should be sparse in some domain. In radar applications, the transmit waveforms are known in advance and the waveform-matched dictionary [25] is often adopted. For a radar with baseband signal $x(t)$ of bandwidth $B/2$, let $\tau_0=1/B$ be its Nyquist sampling interval, and $N=\lceil T/\tau_0\rceil=\lceil BT\rceil$ be the number of Nyquist samples of the receiver signals during the receive time $T$, where $\lceil\cdot\rceil$ denotes the ceiling operation. Then, the waveform-matched dictionary consists of all the time-shifted versions of $s(t)$, i.e., $\{\psi_n(t)|\psi_n(t)=s(t-n\tau_0),n=0,1,\ldots,N-1\}$ at the Nyquist-sampling grids $\{0,\tau_0,\ldots,(N-1)\tau_0\}$. In the waveform-matched dictionary, the time-delay axis is discretized with resolution $\tau_0$. The discretization is reasonable, because the time resolution of the baseband signal $s(t)$ is limited to $1/B$.

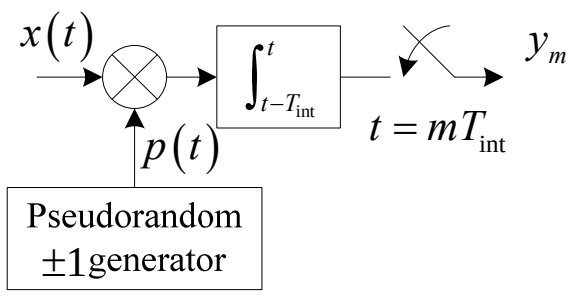

Fig. 2. The structure of RD system.

Assume that the target delays are integral multiples of $\tau_0$, i.e., $t_k\in\{0,\tau_0,\ldots,(N-1)\tau_0\}$. Given the waveform-matched dictionary, $x(t)$ in (4) can be represented as

$$x(t)=\sum_{n=0}^{N-1}\sigma_n\psi_n(t)=\boldsymbol{\psi}(t)\boldsymbol{\sigma},\tag{5}$$

where $\boldsymbol{\sigma}=[\sigma_0,\sigma_1,\ldots,\sigma_{N-1}]^{\mathrm{T}}$ is the sparse coefficient vector to be determined, and $\boldsymbol{\psi}(t)=[\psi_0(t),\psi_1(t),\ldots,\psi_{N-1}(t)]$ is a set of dictionary waveforms. Note that, there are $(N-K)$ zero coefficients in vector $\boldsymbol{\sigma}$ for $K$ targets. As $K\ll N$, $x(t)$ is said to be $K$-sparse signal, where the sparsity level $K$ exactly equals the number of the targets.

### B. Compressive Sampling of Radar Echoes

As introduced in Section I, several AIC systems can be used to implement low-rate sampling of the radar echo signals. The RD AIC system is well developed and has become the fundamental block of RMPI, QuadCS and Xampling systems. For this sake, we take the RD system as an example to illustrate the principle of the compressive sampling in this section.

Fig. 2 shows the basic structure of the RD system. The input analog signal $x(t)$ is first mixed by a random-binary signal

$$p(t)=\pm1,\ t\in\left[k/B_p,\ (k+1)/B_p\right),\ \ k=0,1,2,\ldots,\tag{6}$$

where $B_p\geq B$. Signal $p(t)$ is termed as a chipping sequence with a chipping rate above the Nyquist rate $1/\tau_0$ of the base-band signal. The mixed signal passes through a low-pass filter to prevent aliasing, and the filtered signal is then sampled. For the accumulated low-pass filter, the compressive samples are given by

$$y_m=\int_{(m-1)T_{\mathrm{int}}}^{mT_{\mathrm{int}}}x(t)p(t)dt,\quad m=1,2,\ldots,M,\tag{7}$$

where $T_{\mathrm{int}}=R\tau_0$, and $R>1$ is an integer referred to as the down sampling rate. During the receive time $T$, we can acquire $M=\lfloor T/T_{\mathrm{int}}\rfloor$ low-rate samples, where $\lfloor\cdot\rfloor$ denotes the floor operation.

For the waveform-matched dictionary, letting its elements pass through the RD system yields

$$a_{m,n}=\int_{(m-1)T_{\mathrm{int}}}^{mT_{\mathrm{int}}}\psi_n(t)p(t)dt,\quad m=1,2,\ldots,M.\tag{8}$$

Combining (5), (7) with (8), we have

$$y_m=\sum_{n=0}^{N-1}\sigma_n a_{m,n},\quad m=1,\ldots,M.\tag{9}$$

Define $\mathbf{y}=[y_1,y_2,\ldots,y_M]^T$ and $\mathbf{a}_n=[a_{1,n},a_{2,n},\ldots,a_{M,n}]^T$. Then, (9) can be represented in matrix form as

$$\mathbf{y}=\sum_{n=0}^{N-1}\sigma_n\mathbf{a}_n=\mathbf{A}\boldsymbol{\sigma},\tag{10}$$

where $\mathbf{A}=[\mathbf{a}_0,\mathbf{a}_1,\ldots,\mathbf{a}_{N-1}]$ is an $M\times N$ matrix. Vector $\mathbf{y}$ and matrix $\mathbf{A}$ are referred to as the measurement vector and the measurement matrix, respectively. Because $R>1$, it is clear that $M<N$ and (10) is thus an underdetermined equation.

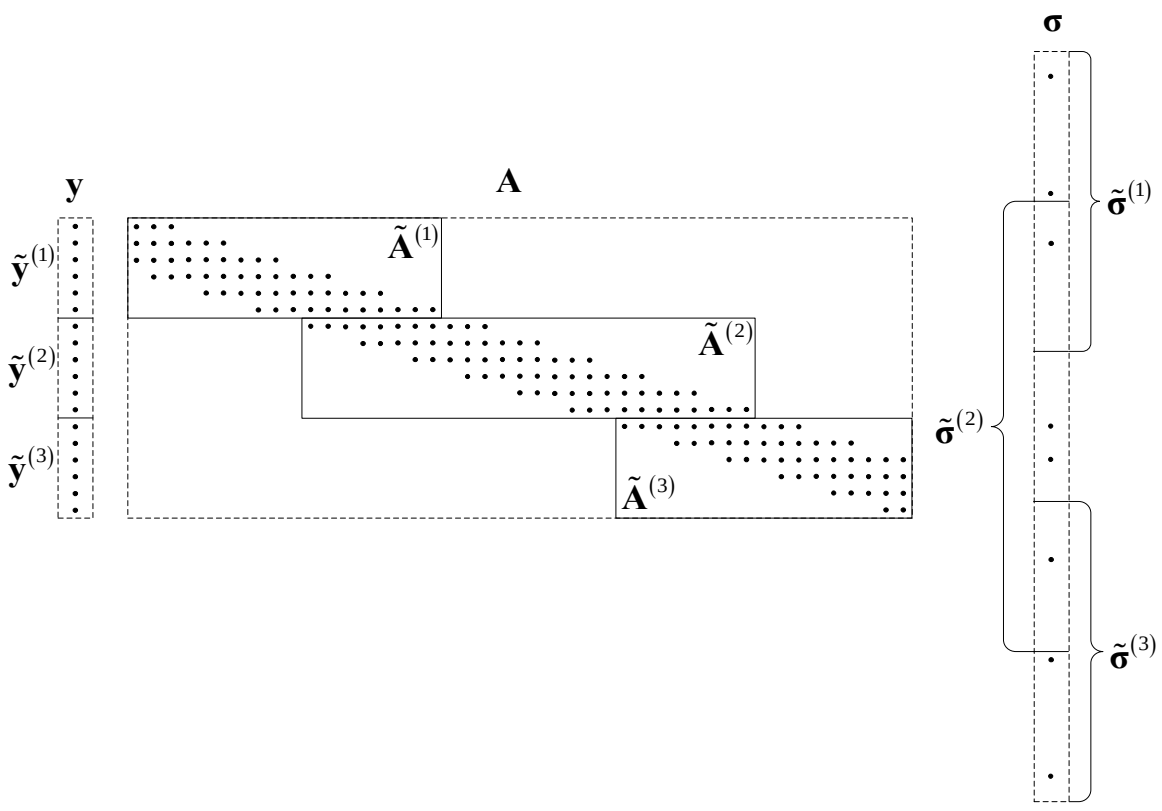

Fig. 3 Schematic illustration of implementing a large-scale reconstruction problem as multiple small-scale reconstruction problems.

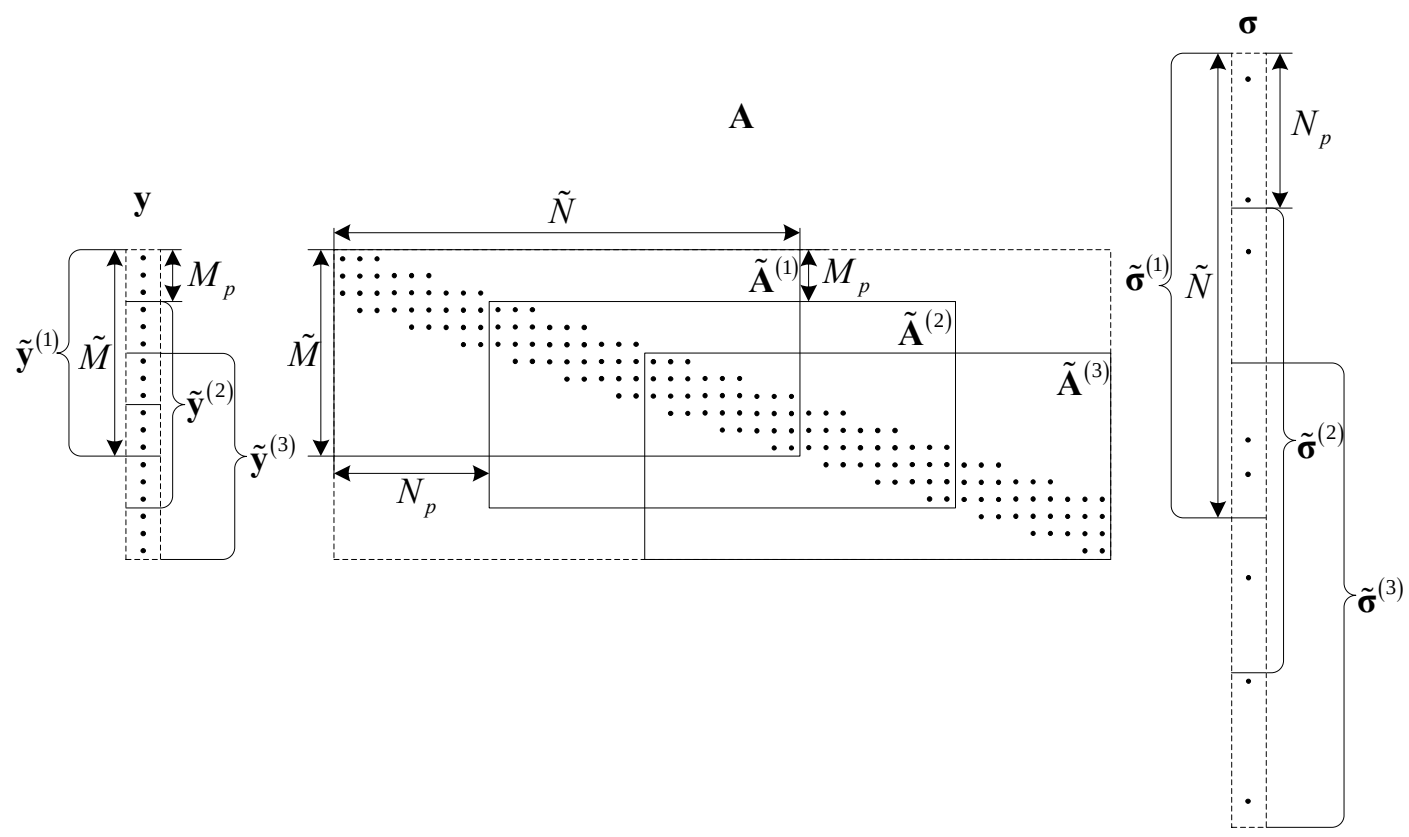

Fig. 4. The schematic illustration of new segmenting.

The reconstruction of the radar echo signal $x(t)$ is equivalent to the recovery of the sparse vector $\boldsymbol{\sigma}$. When matrix $\mathbf{A}$ satisfies the RIP condition (2), the sparse vector $\boldsymbol{\sigma}$ can be exactly reconstructed by solving (1) or (3). As discussed in Section I, the current hardware and computational capacity are infeasible to implement such a large-scale reconstruction. This paper provides an implementable reconstruction technique upon exploiting the structure characteristic of the measurement matrix.

## III. Structure of Measurement Matrix and Sliding Reconstruction

The structure of the measurement matrix results from the combination of RD system and waveform-matched dictionary. Because the transmit radar waveform $s(t)$ has a finite duration of $T_p$, i.e., $s(t)=0$ for $t \notin [0,T_p)$, $\psi_n(t)=s(t-n\tau_0)$ of $\boldsymbol{\psi}(t)$ also has a finite duration, i.e., $\psi_n(t)=0$ for $t \notin [n\tau_0, T_p+n\tau_0)$, $n=0,1,\ldots,N-1$. Then, it is observed from (8) that the $n$-th column of the measurement matrix $\mathbf{A}$ takes nonzero values only in finite indexes, i.e., $a_{m,n}=0$ for $m \le \lfloor n/R \rfloor$ or $m \ge \lceil (T_p+n\tau_0)/T_{\text{int}} \rceil + 1$. That is, the measurement matrix $\mathbf{A}$ is banded matrix with non-zero elements positioned around the diagonal and there are overlaps between the rows. An example with $B=10MHz$, $T_p=0.9\mu s$, $T=5.4\mu s$ and $R=3$ is shown in Fig. 1 in which $M=18$ and $N=45$.

With such a special structure of the matrix $\mathbf{A}$, a straightforward approach, as shown in Fig. 3, is to divide the measurement vector $\mathbf{y}$ into $L$ non-overlapping segments, $\tilde{\mathbf{y}}^{(l)}, l=1,2,\ldots,L$, whereas the sparse coefficient vector $\boldsymbol{\sigma}$ is partitioned into overlapping segments, $\tilde{\boldsymbol{\sigma}}^{(l)}, l=1,2,\ldots,L$, such that a segment contains all the entries that are responsible to a segmented measurement vector. Let $\tilde{\mathbf{A}}^{(l)}$ be the measurement sub-matrix which is formulated by extracting the rows of $\mathbf{A}$ that contribute to $\tilde{\mathbf{y}}^{(l)}$. Then the large-size sparse reconstruction problem (3) can be decomposed into $L$ small-scale ones,

$$\begin{cases} \min\limits_{\tilde{\boldsymbol{\sigma}}^{l}} \left\| \tilde{\boldsymbol{\sigma}}^{(l)} \right\|_1 \\ \text{s.t. } \tilde{\mathbf{y}}^{(l)} = \tilde{\mathbf{A}}^{(l)} \tilde{\boldsymbol{\sigma}}^{(l)}, \end{cases} \tag{11}$$

for $l=1,2,\ldots,L$. As the dimension of (11) can be set in the implementable range, the overall complexity and storage requirement can be significantly reduced. Once we obtain the solution of (11) for $l=1,2,\ldots,L$, we should be able to recover the Nyquist samples of the radar echo signals.

One important problem in (11) is that its feasibility is not guaranteed because the measurement sub-matrix $\tilde{\mathbf{A}}^{(l)}$ does not necessarily satisfy the RIP condition, even if the original problem in (3) does. As we discussed in the previous section, the column vectors $\mathbf{a}_n$ of the measurement matrix $\mathbf{A}$ is obtained by passing $\psi_n(t)$ through the RD system. Then, according to the segmentation of $\mathbf{y}$ and the generation of $\tilde{\mathbf{A}}^{(l)}$, we can find that the last few columns of $\tilde{\mathbf{A}}^{(l)}$ cannot contain all compressive samples of the corresponding elements. There is a high dependency between the last few columns of $\tilde{\mathbf{A}}^{(l)}$, even completely dependent. Therefore, $\tilde{\mathbf{A}}^{(l)}$ cannot satisfy the RIP condition (2). We also notice that there are overlapped estimates between two consecutive estimates $\tilde{\boldsymbol{\sigma}}^{(l-1)}$ and $\tilde{\boldsymbol{\sigma}}^{(l)}$.

In the next section, we develop a new segment-sliding reconstruction scheme which guarantees the RIP condition to be satisfied as long as it is satisfied in the original problem of (3).

## IV. Segment-Sliding Reconstruction Scheme

For simplicity, let the receive time $T$ be $P$ times the radar pulse width $T_p$, i.e., $T=PT_p$ with integer $P \gg 1$. Denote $N_p = N/P$ and $M_p = N/(RP)$ as the number of Nyquist samples and that of the compressive samples in a pulse width, respectively.

Referring to Fig. 4, we modify sub-vector $\tilde{\boldsymbol{\sigma}}^{(l)}$ to include $S$ radar pulses, where $1<S<P$ is an integer, and then have the overlapping sub-vectors $\tilde{\boldsymbol{\sigma}}^{(l)}$ as,

$$\tilde{\boldsymbol{\sigma}}^{(l)} = \boldsymbol{\sigma}\left(\tilde{n} : \tilde{n}+\tilde{N}-1\right), \quad l=1,2,\ldots,L, \tag{12}$$

where $\tilde{n}=(l-1)N_p$, and $\tilde{N}=SN_p$ is the length of the sub-vector $\tilde{\boldsymbol{\sigma}}^{(l)}$. There are $L=P-S$ sparse sub-vectors. Moreover, let us express $\tilde{\boldsymbol{\sigma}}^{(l)}$ as $\tilde{\boldsymbol{\sigma}}^{(l)} = [(\tilde{\boldsymbol{\sigma}}_1^{(l)})^T, (\tilde{\boldsymbol{\sigma}}_2^{(l)})^T, \ldots, (\tilde{\boldsymbol{\sigma}}_S^{(l)})^T]^T$ where $\tilde{\boldsymbol{\sigma}}_s^{(l)} = \tilde{\boldsymbol{\sigma}}^{(l)}\left((s-1)N_p : sN_p-1\right)$ is of length $N_p$. The selection of $S$ depends on the computational capacity and sparsity in the sub-vector $\tilde{\boldsymbol{\sigma}}^{(l)}$. In the following, we assume that sub-vectors $\tilde{\boldsymbol{\sigma}}^{(l)}$ is sparse for all $l=1,2,\ldots,L$.

Similarly, we segment the measurement vector $\mathbf{y}$ into overlapping measurement sub-vectors $\tilde{\mathbf{y}}^{(l)}$,

$$\tilde{\mathbf{y}}^{(l)} = \mathbf{y}\left(\tilde{m} : \tilde{m}+\tilde{M}-1\right), \quad l=1,2,\ldots,L, \tag{13}$$

where $\tilde{m}=(l-1)M_p+1$, and $\tilde{M}=(S+1)M_p$ is the length of the

sub-vector $\tilde{\mathbf{y}}^{(l)}$. Thus the sub-vector $\tilde{\mathbf{y}}^{(l)}$ is downward shifted $M_p$ compressive samples in comparison with $\tilde{\mathbf{y}}^{(l-1)}$.

By extracting the columns and rows of $\mathbf{A}$ corresponding to $\tilde{\boldsymbol{\sigma}}^{(l)}$ and $\tilde{\mathbf{y}}^{(l)}$, we can formulate an $\tilde{M}\times\tilde{N}$ measurement sub-matrix $\tilde{\mathbf{A}}^{(l)}$ as

$$\tilde{\mathbf{A}}^{(l)}=\mathbf{A}\left(\tilde{m}:\tilde{m}+\tilde{M}-1,\tilde{n}:\tilde{n}+\tilde{N}-1\right),\ l=1,2,\ldots,L. \tag{14}$$

Similar to the partition of $\tilde{\boldsymbol{\sigma}}^{(l)}$, we can express $\tilde{\mathbf{A}}^{(l)}$ as $\tilde{\mathbf{A}}^{(l)}=[\tilde{\mathbf{A}}_1^{(l)},\tilde{\mathbf{A}}_2^{(l)},\ldots,\tilde{\mathbf{A}}_S^{(l)}]$ where $\tilde{\mathbf{A}}_s^{(l)}=\tilde{\mathbf{A}}^{(l)}(1:\tilde{M},(s-1)N_p:sN_p-1)$.

In practice, we can downward shift multiple $N_p$ coefficient elements to formulate the sub-vector $\tilde{\boldsymbol{\sigma}}^{(l)}$. Then, the sub-vector $\tilde{\mathbf{y}}^{(l)}$ and the measurement sub-matrix $\tilde{\mathbf{A}}^{(l)}$ are adjusted correspondingly. For simplicity, we assume that $N_p$ coefficient elements are shifted in the theoretical analyses.

As can be observed in Fig. 4, $\tilde{\mathbf{y}}^{(l)}$ is related to three sub-vectors $\tilde{\boldsymbol{\sigma}}_1^{(l-1)}$, $\tilde{\boldsymbol{\sigma}}^{(l)}$, and $\tilde{\boldsymbol{\sigma}}_S^{(l+1)}$. Define

$$\underline{\tilde{\mathbf{A}}}^{(l-1)}=\begin{bmatrix}\tilde{\mathbf{A}}_1^{(l-1)}\left(M_p+1:\tilde{M},0:N_p-1\right)\\ \mathbf{0}_{M_p\times N_p}\end{bmatrix}, \tag{15}$$

$$\underset{\sim}{\tilde{\mathbf{A}}}^{(l+1)}=\begin{bmatrix}\mathbf{0}_{M_p\times N_p}\\ \tilde{\mathbf{A}}_S^{(l+1)}\left(1:\tilde{M}-M_p,0:N_p-1\right)\end{bmatrix}. \tag{16}$$

Then, $\underline{\tilde{\mathbf{A}}}^{(l-1)}\tilde{\boldsymbol{\sigma}}_1^{(l-1)}$ and $\underset{\sim}{\tilde{\mathbf{A}}}^{(l+1)}\tilde{\boldsymbol{\sigma}}_S^{(l+1)}$ are the contributions to $\tilde{\mathbf{y}}^{(l)}$ from $\tilde{\boldsymbol{\sigma}}_1^{(l-1)}$ and $\tilde{\boldsymbol{\sigma}}_S^{(l+1)}$, respectively. The *l*-th measurement sub-vector $\tilde{\mathbf{y}}^{(l)}$ is given by

$$\tilde{\mathbf{y}}^{(l)}=\begin{cases}\tilde{\mathbf{A}}^{(l)}\tilde{\boldsymbol{\sigma}}^{(l)}+\underset{\sim}{\tilde{\mathbf{A}}}^{(l+1)}\tilde{\boldsymbol{\sigma}}_S^{(l+1)}, & l=1,\\ \underline{\tilde{\mathbf{A}}}^{(l-1)}\tilde{\boldsymbol{\sigma}}_1^{(l-1)}+\tilde{\mathbf{A}}^{(l)}\tilde{\boldsymbol{\sigma}}^{(l)}+\underset{\sim}{\tilde{\mathbf{A}}}^{(l+1)}\tilde{\boldsymbol{\sigma}}_S^{(l+1)}, & l=2,3,\ldots,L-1,\\ \underline{\tilde{\mathbf{A}}}^{(l-1)}\tilde{\boldsymbol{\sigma}}_1^{(l-1)}+\tilde{\mathbf{A}}^{(l)}\tilde{\boldsymbol{\sigma}}^{(l)}, & l=L.\end{cases} \tag{17}$$

Define

$$\underline{\tilde{\mathbf{y}}}^{(l)}=\begin{cases}\tilde{\mathbf{y}}^{(l)}, & l=1,\\ \tilde{\mathbf{y}}^{(l)}-\underline{\tilde{\mathbf{A}}}^{(l-1)}\tilde{\boldsymbol{\sigma}}_1^{(l-1)}, & l=2,3,\ldots,L.\end{cases} \tag{18}$$

Then, (17) can be represented as

$$\underline{\tilde{\mathbf{y}}}^{(l)}=\begin{cases}\tilde{\mathbf{A}}^{(l)}\tilde{\boldsymbol{\sigma}}^{(l)}+\underset{\sim}{\tilde{\mathbf{A}}}^{(l+1)}\tilde{\boldsymbol{\sigma}}_S^{(l+1)}, & l=1,2,\ldots,L,\\ \tilde{\mathbf{A}}^{(l)}\tilde{\boldsymbol{\sigma}}^{(l)}, & l=L.\end{cases} \tag{19}$$

Define

$$\underline{\tilde{\mathbf{n}}}^{(l)}=\begin{cases}\underset{\sim}{\tilde{\mathbf{A}}}^{(l+1)}\tilde{\boldsymbol{\sigma}}_S^{(l+1)}, & l=1,2,\ldots,L-1,\\ \mathbf{0}_{M_p\times N_p}, & l=L.\end{cases} \tag{20}$$

We can rewrite (19) in a general form of compressive reconstruction in the presence of noise as

$$\underline{\tilde{\mathbf{y}}}^{(l)}=\tilde{\mathbf{A}}^{(l)}\tilde{\boldsymbol{\sigma}}^{(l)}+\underline{\tilde{\mathbf{n}}}^{(l)},\quad l=1,2,\ldots,L \tag{21}$$

It should be noted that, different from the formulation in Fig. 3, the sub-matrix $\tilde{\mathbf{A}}^{(l)}$ in the above expression satisfies the RIP condition as long as $\mathbf{A}$ does. This is because each column of $\tilde{\mathbf{A}}^{(l)}$ completely contains all nonzero elements of the corresponding column of $\mathbf{A}$.

As revealed by (20), we take the partial measurements in the $(l+1)$-th segment as the noise in the *l*-th segment. Thus, there will be an estimation error when solving (21). Meantime, the model (21) is formulated under the assumption that we have exactly obtained the $\tilde{\boldsymbol{\sigma}}^{(l-1)}$. Therefore, any estimation inaccuracy in $\tilde{\boldsymbol{\sigma}}^{(l-1)}$ will affect the estimate of $\tilde{\boldsymbol{\sigma}}^{(l)}$. In the next section, we will conduct the analyses of these interferences on the estimation performance.

Considering an inaccurate estimate of $\tilde{\boldsymbol{\sigma}}^{(l-1)}$, we refine the model (21). Let $\hat{\tilde{\boldsymbol{\sigma}}}_1^{(l-1)}$ be the estimate of $\tilde{\boldsymbol{\sigma}}_1^{(l-1)}$ and $\Delta\tilde{\boldsymbol{\sigma}}_1^{(l-1)}$ be the error between $\tilde{\boldsymbol{\sigma}}_1^{(l-1)}$ and $\hat{\tilde{\boldsymbol{\sigma}}}_1^{(l-1)}$, i.e., $\Delta\tilde{\boldsymbol{\sigma}}_1^{(l-1)}=\tilde{\boldsymbol{\sigma}}_1^{(l-1)}-\hat{\tilde{\boldsymbol{\sigma}}}_1^{(l-1)}$. Then, (17) becomes

$$\tilde{\mathbf{y}}^{(l)}=\begin{cases}\tilde{\mathbf{A}}^{(l)}\tilde{\boldsymbol{\sigma}}^{(l)}+\underset{\sim}{\tilde{\mathbf{A}}}^{(l+1)}\tilde{\boldsymbol{\sigma}}_S^{(l+1)}, & l=1,\\ \underline{\tilde{\mathbf{A}}}^{(l-1)}\hat{\tilde{\boldsymbol{\sigma}}}_1^{(l-1)}+\tilde{\mathbf{A}}^{(l)}\tilde{\boldsymbol{\sigma}}^{(l)}+\underline{\tilde{\mathbf{A}}}^{(l-1)}\Delta\tilde{\boldsymbol{\sigma}}_1^{(l-1)}+\underset{\sim}{\tilde{\mathbf{A}}}^{(l+1)}\tilde{\boldsymbol{\sigma}}_S^{(l+1)}, & l=2,3,\ldots,L-1,\\ \underline{\tilde{\mathbf{A}}}^{(l-1)}\hat{\tilde{\boldsymbol{\sigma}}}_1^{(l-1)}+\tilde{\mathbf{A}}^{(l)}\tilde{\boldsymbol{\sigma}}^{(l)}+\underline{\tilde{\mathbf{A}}}^{(l-1)}\Delta\tilde{\boldsymbol{\sigma}}_1^{(l-1)}, & l=L.\end{cases} \tag{22}$$

Define

$$\underline{\tilde{\mathbf{y}}}^{(l)}=\begin{cases}\tilde{\mathbf{y}}^{(l)}, & l=1,\\ \tilde{\mathbf{y}}^{(l)}-\underline{\tilde{\mathbf{A}}}^{(l-1)}\hat{\tilde{\boldsymbol{\sigma}}}_1^{(l-1)}, & l=2,3,\ldots,L,\end{cases} \tag{23}$$

$$\underline{\tilde{\mathbf{n}}}^{(l)}=\begin{cases}\underset{\sim}{\tilde{\mathbf{A}}}^{(l+1)}\tilde{\boldsymbol{\sigma}}_S^{(l+1)}, & l=1,\\ \underline{\tilde{\mathbf{A}}}^{(l-1)}\Delta\tilde{\boldsymbol{\sigma}}_1^{(l-1)}+\underset{\sim}{\tilde{\mathbf{A}}}^{(l+1)}\tilde{\boldsymbol{\sigma}}_S^{(l+1)}, & l=2,3,\ldots,L-1,\\ \underline{\tilde{\mathbf{A}}}^{(l-1)}\Delta\tilde{\boldsymbol{\sigma}}_1^{(l-1)}, & l=L.\end{cases} \tag{24}$$

Then, we have

$$\underline{\tilde{\mathbf{y}}}^{(l)}=\tilde{\mathbf{A}}^{(l)}\tilde{\boldsymbol{\sigma}}^{(l)}+\underline{\tilde{\mathbf{n}}}^{(l)},\ \ l=1,2,\ldots,L. \tag{25}$$

When $\tilde{\boldsymbol{\sigma}}^{(l)}$ is sparse and $\left\|\underline{\tilde{\mathbf{n}}}^{(l)}\right\|_2\le\eta^{(l)}$, we can obtain the estimate $\hat{\tilde{\boldsymbol{\sigma}}}^{(l)}$ of $\tilde{\boldsymbol{\sigma}}^{(l)}$ by solving the following constrained optimization problem:

$$\begin{cases}\min\left\|\tilde{\boldsymbol{\sigma}}^{(l)}\right\|_1\\ \text{s.t.}\ \left\|\underline{\tilde{\mathbf{y}}}^{(l)}-\tilde{\mathbf{A}}^{(l)}\tilde{\boldsymbol{\sigma}}^{(l)}\right\|_2\le\eta^{(l)}.\end{cases} \tag{26}$$

The "noise" $\underline{\tilde{\mathbf{n}}}^{(l)}$ is generated during estimation process that integrates the interferences from the estimation error in the previous segment and the partial measurement in the subsequent segment. For convenience, we refer to $\underline{\tilde{\mathbf{n}}}^{(l)}$ as the *virtual noise* sub-vector. Similarly, we name $\underline{\tilde{\mathbf{y}}}^{(l)}$ as the *virtual measurement* sub-vector. It is obvious from (24) that the virtual noise $\underline{\tilde{\mathbf{n}}}^{(l)}$ consists of $\underline{\tilde{\mathbf{A}}}^{(l-1)}\Delta\tilde{\boldsymbol{\sigma}}_1^{(l-1)}$ and $\underset{\sim}{\tilde{\mathbf{A}}}^{(l+1)}\tilde{\boldsymbol{\sigma}}_S^{(l+1)}$ generated by the inaccurate estimate in segment $l-1$ and the partial measurement in segment $l+1$. For convenience, we call them forward and backward virtual noises, respectively.

The virtual noise $\underline{\tilde{\mathbf{n}}}^{(l)}$ is closely related with measurement matrix. From the definitions of $\underline{\tilde{\mathbf{A}}}^{(l-1)}$ and $\underset{\sim}{\tilde{\mathbf{A}}}^{(l+1)}$, respectively depicted in (15) and (16), we observe that the forward (backward) virtual noise exists only in the range of $[1:M_p]$ ( $[\tilde{M}-M_p+1:\tilde{M}]$ ). As discussed in the next section, the backward virtual noise level is in general much larger than that of the forward virtual noise.

It is seen from (25) that we can formulate a series of sparse reconstructions by properly segmenting the measurement vector and the coefficient vector. Then a large-scale reconstruction problem is decomposed into a series of small-scale ones. We call the method as segment-sliding sparse reconstruction (SegSR). Algorithm 1 describes the SegSR process in detail.

Solving (26) is one of the dominant steps in the implementation of the proposed technique. Theoretically, any sparse estimation algorithm can be used. However, the formulation (25) has three distinct characteristics: (1) The sparsity level of each segment is in general unknown; (2) There are some overlaps between two consecutive segments of the sparse vector $\boldsymbol{\sigma}$; (3) The distributions of the virtual noise are different for each virtual measurements. These characteristics can be incorporated into the estimate of $\tilde{\boldsymbol{\sigma}}^{(l)}$ to improve the algorithm

**Algorithm 1. SegSR Scheme**

**Input:** $S$, $M_p$, $N_p$, $\tilde{M}$, $\tilde{N}$ and $L$

**Output:** Estimated sparse vector $\hat{\boldsymbol{\sigma}}$

**Steps:**

1) Initialize $l=1$.
2) Extract the $\tilde{M}\times 1$ measurement sub-vector $\tilde{\mathbf{y}}^{(l)}$ and the $\tilde{M}\times\tilde{N}$ measurement sub-matrix $\tilde{\mathbf{A}}^{(l)}$.
3) Calculate the virtual measurement sub-vector $\underline{\tilde{\mathbf{y}}}^{(l)}$ by (23).
4) Solve (26) and obtain the estimate $\hat{\underline{\tilde{\boldsymbol{\sigma}}}}^{(l)}$ of $\underline{\tilde{\boldsymbol{\sigma}}}^{(l)}$.
5) Let $l=l+1$. If $l>L$, go to Step 6); otherwise, generate the $\tilde{M}\times\tilde{N}$ sub-matrix $\underline{\tilde{\mathbf{A}}}^{(l-1)}$ by (15), extract the sub-vector $\hat{\tilde{\boldsymbol{\sigma}}}_1^{(l-1)}$ from the estimated $\hat{\underline{\tilde{\boldsymbol{\sigma}}}}^{(l-1)}$ and go to Step 2).
6) Formulate the estimated sparse vector $\hat{\boldsymbol{\sigma}}$ of $\boldsymbol{\sigma}$ as
$$\begin{cases}\hat{\boldsymbol{\sigma}}\left((l-1)N_p : lN_p-1\right)=\hat{\tilde{\boldsymbol{\sigma}}}_1^{(l)}, & l=1,\ldots,L-1,\\ \hat{\boldsymbol{\sigma}}\left((l-1)N_p : N-1\right)=\hat{\tilde{\boldsymbol{\sigma}}}^{(L)}, & l=L.\end{cases}$$

efficiency. Toward this objective, we make some modifications on the OMP algorithm with partially known support (OMP-PKS) [31] to solve (26). Algorithm 2 presents the computational flowchart of the algorithm, termed as the two-step OMP-process (TOMPP). TOMPP consists of three main processes. One is the initialization to take into account the partially known support from the previous segment. The second one is the OMP process with a threshold $\zeta_1$ to obtain a rough estimate (Steps 2~4) derived for the whole virtual noise. The last process (Steps 5~7) is the refinement OMP process with threshold $\zeta_2$ for the sparse estimate in sparse range $[0:\tilde{N}-N_p-1]$ because of the virtual noise structure. The effectiveness of the TOMPP is validated by theoretical analyses and simulation experiments in the next three sections.

**Algorithm 2. TOMPP**

**Input:** $\underline{\tilde{\mathbf{y}}}^{(l)}$, $\tilde{\mathbf{A}}^{(l)}$, $N_p$, $\tilde{N}$, $\hat{\tilde{\boldsymbol{\sigma}}}^{(l-1)}$, threshold parameters $\zeta_1$ and $\zeta_2$ ($\zeta_2<\zeta_1$)

**Output:** Estimated sparse sub-vector $\hat{\tilde{\boldsymbol{\sigma}}}^{(l)}$

**Steps:**

1) Initialize the index set $\Lambda^{[0]}$ as the known support in the $l$-th segment estimate, $\Lambda^{[0]}=\text{support}\left(\hat{\tilde{\boldsymbol{\sigma}}}^{(l-1)}\left(N_p:\tilde{N}-1\right)\right)$; the residual $\tilde{\mathbf{r}}^{[0]}=\left(\mathbf{I}-\mathbf{P}_0\right)\underline{\tilde{\mathbf{y}}}^{(l)}$, where $\mathbf{P}_0=\tilde{\mathbf{A}}^{(l)}_{\Lambda^{[0]}}\left(\tilde{\mathbf{A}}^{(l)}_{\Lambda^{[0]}}\right)^{\dagger}$ denotes the projection onto the linear space spanned by the columns of $\tilde{\mathbf{A}}^{(l)}_{\Lambda^{[0]}}$. Let $i=1$.
2) Find the column of $\tilde{\mathbf{A}}^{(l)}$ that is most correlated with the residual $\tilde{\mathbf{r}}^{[i-1]}$, i.e.,
$$\lambda^{[i]}=\underset{j=0,1,\ldots,\tilde{N}-1}{\arg\max}\left|\left\langle\tilde{\mathbf{a}}_j^{(l)},\tilde{\mathbf{r}}^{[i-1]}\right\rangle\right|,$$
and update the index set $\Lambda^{[i]}=\Lambda^{[i-1]}\bigcup\left\{\lambda^{[i]}\right\}$.
3) Update the residual $\tilde{\mathbf{r}}^{[i]}=\left(\mathbf{I}-\mathbf{P}_i\right)\underline{\tilde{\mathbf{y}}}^{(l)}$.
4) If $\left\|\tilde{\mathbf{r}}^{[i-1]}\right\|_2-\left\|\tilde{\mathbf{r}}^{[i]}\right\|_2\le\zeta_1$, let $i=i+1$ and go to Step 5); otherwise, let $i=i+1$ and return to Step 2).
5) For columns $0\sim\tilde{N}-N_p-1$ of $\tilde{\mathbf{A}}^{(l)}$, find the column that is most correlated with the residual $\tilde{\mathbf{r}}^{[i-1]}$, i.e.,
$$\lambda^{[i]}=\underset{j=0,1,\ldots,\tilde{N}-N_p-1}{\arg\max}\left|\left\langle\tilde{\mathbf{a}}_j^{(l)},\tilde{\mathbf{r}}^{[i-1]}\right\rangle\right|,$$
and update index set $\Lambda^{[i]}=\Lambda^{[i-1]}\bigcup\left\{\lambda^{[i]}\right\}$.
6) Update the residual $\tilde{\mathbf{r}}^{[i]}=\left(\mathbf{I}-\mathbf{P}_i\right)\underline{\tilde{\mathbf{y}}}^{(l)}$.
7) If $\left\|\tilde{\mathbf{r}}^{[i-1]}\right\|_2-\left\|\tilde{\mathbf{r}}^{[i]}\right\|_2\le\zeta_2$, go to Step 8); otherwise, let $i=i+1$ and return to Step 5).
8) Calculate the estimate $\hat{\tilde{\boldsymbol{\sigma}}}^{(l)}$ by $\hat{\tilde{\boldsymbol{\sigma}}}^{(l)}_{\Lambda^{[i]}}=\left(\tilde{\mathbf{A}}^{(l)}_{\Lambda^{[i]}}\right)^{\dagger}\underline{\tilde{\mathbf{y}}}^{(l)}$.

## V. Effects of Errors on Reconstruction Performance

From the SegSR flowchart, we find that the estimate in the current segment will affect the estimate of the subsequent segment. The estimation performance in each segment will be affected by the virtual noise. This section takes the OMP algorithm as an example to analyze the effect of the virtual noise on the estimation performance in terms of the sparse entry positions and sparse coefficients.

### *A. Effect of the Virtual Noise on Position Estimation*

OMP is an iterative algorithm, which selects at each step the column of the measurement matrix which is most highly correlated with the current residuals as an estimate of one of sparse positions. References [32]-[34] have conducted thorough analyses on the recovery performance of sparse positions in $\ell_2$-bounded noise and $\ell_\infty$-bounded noise. These analyses indicate that the sparse positions can be estimated exactly by OMP for the non-zero sparse element when its magnitude is greater than some value. For the underlying problem considered in this paper, we first show that the virtual noise (24) is $\ell_2$-bounded and $\ell_\infty$-bounded. Then we derive the conditions for correct sparse position estimation.

Before proceeding, we define a few notations. Let $\Gamma^{(l)}$ and $\Gamma_s^{(l)}$ $(s=1,\ldots,S)$ denote the support sets of $\tilde{\boldsymbol{\sigma}}^{(l)}$ and $\tilde{\boldsymbol{\sigma}}_s^{(l)}$ $(s=1,\ldots,S)$, respectively. Let $\Delta\Gamma_1^{(l-1)}$ denote the support sets of $\Delta\tilde{\boldsymbol{\sigma}}_1^{(l-1)}$. Let $\tilde{\delta}_k^{(l)}$ be the $k$-th order RIP constant of matrix $\tilde{\mathbf{A}}^{(l)}$. For the virtual noise in (26), we have:

**Theorem 1**: For $l=1,2,\ldots,L$, the virtual noise $\underline{\tilde{\mathbf{n}}}^{(l)}$ satisfies
$$\left\|\underline{\tilde{\mathbf{n}}}^{(l)}\right\|_2\le\varepsilon_2^{(l)}=\begin{cases}b_2^{(l+1)}, & l=1,\\ a_2^{(l-1)}+b_2^{(l+1)}, & l=2,3,\ldots,L-1,\\ a_2^{(l-1)}, & l=L.\end{cases}\tag{27}$$
$$\left\|\left(\tilde{\mathbf{A}}^{(l)}\right)^H\underline{\tilde{\mathbf{n}}}^{(l)}\right\|_\infty\le\varepsilon_\infty^{(l)}=\begin{cases}b_\infty^{(l+1)}, & l=1,\\ a_\infty^{(l-1)}+b_\infty^{(l+1)}, & l=2,3,\ldots,L-1,\\ a_\infty^{(l-1)}, & l=L.\end{cases}\tag{28}$$
where $a_2^{(l-1)}=\sqrt{1+\tilde{\delta}^{(l-1)}_{\left|\Delta\Gamma_1^{(l-1)}\right|}}\left\|\Delta\tilde{\boldsymbol{\sigma}}_1^{(l-1)}\right\|_2$ and $a_\infty^{(l-1)}=\tilde{\delta}^{(l-1)}_{\left|\Gamma_1^{(l-1)}\right|+1}\left\|\Delta\tilde{\boldsymbol{\sigma}}_1^{(l-1)}\right\|_2$ for $l=2,3,\ldots,L$, $b_2^{(l+1)}=\sqrt{1+\tilde{\delta}^{(l+1)}_{\left|\Gamma_S^{(l+1)}\right|}}\left\|\tilde{\boldsymbol{\sigma}}_S^{(l+1)}\right\|_2$ and $b_\infty^{(l+1)}=\tilde{\delta}^{(l+1)}_{\left|\Gamma_S^{(l+1)}\right|+1}\left\|\tilde{\boldsymbol{\sigma}}_S^{(l+1)}\right\|_2$ for $l=1,2,\ldots,L-1$.

***Proof***: See Appendix A.

From (27) and (28), it is found that the virtual noise $\underline{\tilde{\mathbf{n}}}^{(l)}$ is both $\ell_2$-bounded and $\ell_\infty$-bounded and is limited by $\varepsilon_2^{(l)}$ and $\varepsilon_\infty^{(l)}$, respectively. The upper limits consist of two parts, $a_2^{(l-1)}$ ($a_\infty^{(l-1)}$) and $b_2^{(l+1)}$ ($b_\infty^{(l+1)}$). which are related to the estimate error $\Delta\tilde{\boldsymbol{\sigma}}_1^{(l-1)}$ in the previous segment and the sparse sub-vector $\tilde{\boldsymbol{\sigma}}_S^{(l+1)}$ in the subsequent segment, respectively. The noise limits $\varepsilon_2^{(l)}$ and $\varepsilon_\infty^{(l)}$ are usually dominated by $b_2^{(l+1)}$ and $b_\infty^{(l+1)}$, respectively.

For the $\ell_2$-bounded and $\ell_\infty$-bounded virtual noise, we have the following theorems to find the correct sparse positions.

**Theorem 2:** Suppose that $\left\|\underline{\tilde{\mathbf{n}}}^{(l)}\right\|_2 \le \varepsilon_2^{(l)}$ and $\tilde{\mathbf{A}}^{(l)}$ satisfies condition $\tilde{\delta}^{(l)}_{|\Gamma^{(l)}|+1} < 1\Big/\left(\sqrt{|\Gamma^{(l)}|}+1\right)$. Then OMP with the stopping rule $\left\|\tilde{\mathbf{r}}^{[i]}\right\|_2 \le \varepsilon_2^{(l)}$ will exactly recover the support $\Gamma^{(l)}$ of $\tilde{\boldsymbol{\sigma}}^{(l)}$, if all the nonzero coefficients $\tilde{\sigma}_j^{(l)}$ ( $j \in \Gamma^{(l)}$ ) satisfy

$$\left|\tilde{\sigma}_j^{(l)}\right| \ge \frac{\left(\sqrt{1+\tilde{\delta}^{(l)}_{|\Gamma^{(l)}|+1}}+1\right)\varepsilon_2^{(l)}}{1-\left(\sqrt{|\Gamma^{(l)}|}+1\right)\tilde{\delta}^{(l)}_{|\Gamma^{(l)}|+1}}. \tag{29}$$

**Theorem 3:** Suppose that $\left\|\left(\tilde{\mathbf{A}}^{(l)}\right)^H \underline{\tilde{\mathbf{n}}}^{(l)}\right\|_\infty \le \varepsilon_\infty^{(l)}$ and $\tilde{\mathbf{A}}^{(l)}$ satisfies condition $\tilde{\delta}^{(l)}_{|\Gamma^{(l)}|+1} < 1\Big/\left(\sqrt{|\Gamma^{(l)}|}+1\right)$. Then OMP with the stopping rule $\left\|\left(\tilde{\mathbf{A}}^{(l)}\right)^H \tilde{\mathbf{r}}^{[i]}\right\|_2 \le \varepsilon_\infty^{(l)}$ will exactly recover the support $\Gamma^{(l)}$ of $\tilde{\boldsymbol{\sigma}}^{(l)}$, if all the nonzero coefficients $\tilde{\sigma}_j^{(l)}$ ( $j \in \Gamma^{(l)}$ ) satisfy

$$\left|\tilde{\sigma}_j^{(l)}\right| \ge \frac{\left(\sqrt{1+\tilde{\delta}^{(l)}_{|\Gamma^{(l)}|+1}}+1\right)\sqrt{|\Gamma^{(l)}|}\,\varepsilon_\infty^{(l)}}{1-\left(\sqrt{|\Gamma^{(l)}|}+1\right)\tilde{\delta}^{(l)}_{|\Gamma^{(l)}|+1}}. \tag{30}$$

The proofs of two theorems closely follow those in [33] and thus are omitted here.

Equipped with these two theorems, we can find the correct sparse entry positions by the OMP algorithm if the nonzero coefficients satisfy (29) or (30). However, it should be noted that the two conditions are derived with overall effects of the virtual noise and the resulting minimum nonzero magnitude may be high. In this case, we may miss some nonzero elements of the sparse vector.

As discussed in the previous section, the virtual noise $\underline{\tilde{\mathbf{n}}}^{(l)}$ exists only in the ranges $[1:M_p]$ and $[\tilde{M}-M_p+1:\tilde{M}]$. Then the columns of $\tilde{\mathbf{A}}_s^{(l)}$ $(s=2,\ldots,S)$ are orthogonal to $\underline{\tilde{\mathbf{A}}}^{(l-1)}\Delta\tilde{\boldsymbol{\sigma}}_1^{(l-1)}$ and the columns of $\tilde{\mathbf{A}}_s^{(l)}$ $(s=1,\ldots,S-1)$ are orthogonal to $\underset{\sim}{\tilde{\mathbf{A}}}^{(l+1)}\tilde{\boldsymbol{\sigma}}_S^{(l+1)}$. In particular, when $S \ge 3$, the columns of $\tilde{\mathbf{A}}_s^{(l)}$ $(s=2,\ldots,S-1)$ are orthogonal to both forward virtual noise $\underline{\tilde{\mathbf{A}}}^{(l-1)}\Delta\tilde{\boldsymbol{\sigma}}_1^{(l-1)}$ and backward virtual noise $\underset{\sim}{\tilde{\mathbf{A}}}^{(l+1)}\tilde{\boldsymbol{\sigma}}_S^{(l+1)}$. By the principle of the OMP, the estimation of the sparse positions of $\tilde{\boldsymbol{\sigma}}_1^{(l)}$ is mainly affected by the $\underline{\tilde{\mathbf{A}}}^{(l-1)}\Delta\tilde{\boldsymbol{\sigma}}_1^{(l-1)}$ and, similarly, that of $\tilde{\boldsymbol{\sigma}}_S^{(l)}$ is mainly affected by the $\underset{\sim}{\tilde{\mathbf{A}}}^{(l+1)}\tilde{\boldsymbol{\sigma}}_S^{(l+1)}$. In general, the noise level of $\underset{\sim}{\tilde{\mathbf{A}}}^{(l+1)}\tilde{\boldsymbol{\sigma}}_S^{(l+1)}$ is higher than that of $\underline{\tilde{\mathbf{A}}}^{(l-1)}\Delta\tilde{\boldsymbol{\sigma}}_1^{(l-1)}$. Then we can use refinement OMP process in Algorithm 2 with a small threshold $\zeta_2$ to improve the estimate of the sparse vector $\tilde{\boldsymbol{\sigma}}^{(l)}$ in the range $[0:\tilde{N}-N_p-1]$.

## *B. Effect of the Virtual Noise on Amplitude Estimation*

We now analyze the effect of virtual noise on amplitude estimation of sub-vectors $\tilde{\boldsymbol{\sigma}}_s^{(l)}$ $(s=1,2,\ldots,S)$ of $\tilde{\boldsymbol{\sigma}}^{(l)}$. To simplify the analysis, we assume that the support $\Gamma^{(l)}$ of $\tilde{\boldsymbol{\sigma}}^{(l)}$ is correctly recovered.

With the support $\Gamma^{(l)}$ of $\tilde{\boldsymbol{\sigma}}^{(l)}$, we can obtain the estimate $\hat{\tilde{\boldsymbol{\sigma}}}^{(l)}_{\Gamma^{(l)}}$ of $\tilde{\boldsymbol{\sigma}}^{(l)}_{\Gamma^{(l)}}$ through the least-square solution of

$$\underline{\tilde{\mathbf{y}}}^{(l)} = \tilde{\mathbf{A}}^{(l)}_{\Gamma^{(l)}}\tilde{\boldsymbol{\sigma}}^{(l)}_{\Gamma^{(l)}} + \underline{\tilde{\mathbf{n}}}^{(l)} = \sum\nolimits_{s=1}^{S}\tilde{\mathbf{A}}^{(l)}_{\Gamma_s^{(l)}}\tilde{\boldsymbol{\sigma}}^{(l)}_{\Gamma_s^{(l)}} + \underline{\tilde{\mathbf{n}}}^{(l)}, \quad l=1,2,\ldots,L, \tag{31}$$

where

$$\underline{\tilde{\mathbf{n}}}^{(l)} = \begin{cases} \underset{\sim}{\tilde{\mathbf{A}}}^{(l+1)}_{\Gamma_S^{(l+1)}}\tilde{\boldsymbol{\sigma}}^{(l+1)}_{\Gamma_S^{(l+1)}}, & l=1, \\ \underline{\tilde{\mathbf{A}}}^{(l-1)}_{\Gamma_1^{(l-1)}}\Delta\tilde{\boldsymbol{\sigma}}^{(l-1)}_{\Gamma_1^{(l-1)}} + \underset{\sim}{\tilde{\mathbf{A}}}^{(l+1)}_{\Gamma_S^{(l+1)}}\tilde{\boldsymbol{\sigma}}^{(l+1)}_{\Gamma_S^{(l+1)}}, & l=2,3,\ldots,L-1, \\ \underline{\tilde{\mathbf{A}}}^{(l-1)}_{\Gamma_1^{(l-1)}}\Delta\tilde{\boldsymbol{\sigma}}^{(l-1)}_{\Gamma_1^{(l-1)}}, & l=L. \end{cases} \tag{32}$$

The estimated $\hat{\tilde{\boldsymbol{\sigma}}}^{(l)}_{\Gamma^{(l)}}$ is given by

$$\hat{\tilde{\boldsymbol{\sigma}}}^{(l)}_{\Gamma^{(l)}} = \left(\tilde{\mathbf{A}}^{(l)}_{\Gamma^{(l)}}\right)^{\dagger}\underline{\tilde{\mathbf{y}}}^{(l)}. \tag{33}$$

For $\hat{\tilde{\boldsymbol{\sigma}}}^{(l)}_{\Gamma^{(l)}}$, we have the estimation error bound as in Theorem 4.

**Theorem 4:** For $l=1,2,\ldots,L$, the least-squares estimate error of (31), $\left\|\Delta\tilde{\boldsymbol{\sigma}}^{(l)}_{\Gamma_s^{(l)}}\right\|_2 = \left\|\tilde{\boldsymbol{\sigma}}^{(l)}_{\Gamma_s^{(l)}} - \hat{\tilde{\boldsymbol{\sigma}}}^{(l)}_{\Gamma_s^{(l)}}\right\|_2$ $(s=1,\ldots,S)$, is upper bounded by

$$\left\|\Delta\tilde{\boldsymbol{\sigma}}^{(l)}_{\Gamma_s^{(l)}}\right\|_2 \le \begin{cases} \alpha^{S-s+1}\left\|\tilde{\boldsymbol{\sigma}}^{(l+1)}_{\Gamma_S^{(l+1)}}\right\|_2, & l=1, \\ \beta\left\|\Delta\tilde{\boldsymbol{\sigma}}^{(l-1)}_{\Gamma_1^{(l-1)}}\right\|_2 + \alpha^{S-s+1}\left\|\tilde{\boldsymbol{\sigma}}^{(l+1)}_{\Gamma_S^{(l+1)}}\right\|_2, & l=2,3,\ldots,L-1, \\ \beta\left\|\Delta\tilde{\boldsymbol{\sigma}}^{(l-1)}_{\Gamma_1^{(l-1)}}\right\|_2, & l=L. \end{cases} \tag{34}$$

where $\beta = \alpha^s\left(1-\alpha^{2(S-s+1)}\right)\big/\left(1-\alpha^2\right)$ and $\alpha = \bar{\delta}^{(l)}_{\bar{K}^{(l)}}\big/\left(1-\bar{\delta}^{(l)}_{\bar{K}^{(l)}}\right)$, with $\bar{K}^{(l)}$ and $\bar{\delta}^{(l)}_k$ defined by $\bar{K}^{(l)} = \max\left\{\left|\Gamma^{(l-1)}\right|, \left|\Gamma^{(l)}\right|, \left|\Gamma^{(l+1)}\right|\right\}$ and $\bar{\delta}^{(l)}_k = \max\left\{\tilde{\delta}^{(l-1)}_k, \tilde{\delta}^{(l)}_k, \tilde{\delta}^{(l+1)}_k\right\}$.

***Proof***: See Appendix B.

The error bound (34) not only gives the least-squares bound but also shows the bound distribution in each segment. Note that, when $\bar{\delta}^{(l)}_{\bar{K}^{(l)}} < 1/3$, we have $\alpha < 1/2$ and $\beta < 1$. From the relation $\beta = \alpha^s\left(1-\alpha^{2(S-s+1)}\right)\big/\left(1-\alpha^2\right)$, we know that $\beta$ decreases as $s$ increases. Therefore, the estimation error due to the forward virtual noise decreases as $s$ increases. However, the error introduced by the backward virtual noise increases with $s$ because of the $\alpha^{S-s+1}$ term. Similar to the analyses on the noise bounds, the error bound of the current segment is usually dominated by the error introduced in the subsequent segment. With the effect behavior of the subsequent segment, we can deduce that the sub-vector $\tilde{\boldsymbol{\sigma}}_1^{(l)}$ of the sparse vector $\tilde{\boldsymbol{\sigma}}^{(l)}$ can be more accurately estimated than other sub-vectors $\tilde{\boldsymbol{\sigma}}_s^{(l)}$ $(s=2,\ldots,S)$. Therefore, we downward slide one pulse width in Algorithm 1. While we can downward slide by multiple pulse widths. the estimation error will nevertheless increase, as the simulated results will show in Section VII.

# VI. Storage Requirements and Computational Complexity

## *A. Storage Requirements*

For the reconstruction problem (3), we need to store $M \times N$ matrix $\mathbf{A}$ which occupies $8MN = 8P(P-1)M_pN_p$ bytes using the standard IEEE double precision. For the SegSR scheme, we only need to store $\tilde{M} \times \tilde{N}$ sub-matrix $\tilde{\mathbf{A}}^{(l)}$ involved in sparse reconstruction sub-problem (26). Therefore, the required storage space is $8\tilde{M}\tilde{N} = 8S(S+1)M_pN_p$ bytes. Table I shows the storage requirements with $P = 249$, $M_p = 200$ and $N_p = 1000$. It is seen that the storage space is significantly reduced when the SegSR scheme is used.

TABLE I
STORAGE REQUIREMENTS

| Reconstruction by (3) | SegSR scheme | | |
|---|---|---|---|
| | $S=2$ | $S=3$ | $S=4$ |
| 92.02*GB* | 9.16*MB* | 18.31*MB* | 30.52*MB* |

### B. *Computational Complexity*

As is well known, the computational complexity for OMP to solve the problem (3) is $O(KMN)$. For the SegSR scheme, the sub-problem (26) is solved by OMP with a complexity of $O\left(\left|\Gamma^{(l)}\right|\tilde{M}\tilde{N}\right)$. Similar to the analysis of storage capacity, the complexity of sub-problem is significantly lower than that of original problem. Therefore, the real-time capacity for SegSR is improved greatly. Furthermore, when we use OMP with partially known support to solve the sub-problem, the real-time capacity can be improved further.

## VII. Simulation Results

In this section, we evaluate the performance of the proposed SegSR scheme through several simulation experiments. Without special statements, 500 realizations are conducted and the averaged results are presented.

We take the following linear frequency modulated (LFM) pulse signal as an example,

$$s(t) = \mathrm{rect}\left(\left(t - T_p/2\right)/T_p\right)\cos\left(\gamma\pi\left(t - T_p/2\right)^2\right), \qquad (35)$$

where $T_p$ is the pulse width, $\gamma = B/T_p$ is the chirp rate, $B$ is the bandwidth of the LFM signal, and $\mathrm{rect}(t/T_p)$ represents a rectangular pulse

$$\mathrm{rect}(t/T_p) = \begin{cases} 1, & -T_p/2 \le t \le T_p/2, \\ 0, & \text{elsewhere.} \end{cases} \qquad (36)$$

In order to make comparison with direct reconstruction, we set the signal parameters such that the measurement size is moderate. In simulations, $B = 100MHz$, $T_p = 10\mu s$ and $T = 100\mu s$. For the RD system, the chipping rate of $p(t)$ is set as $B_p = 100MHz$, the integration time is set as $T_{\text{int}} = 0.05\mu s$, which means that the sampling rate of ADC is $20MHz$ and the down sampling rate is $R = 5$. With these parameters, we can obtain $N = 10000$ Nyquist samples and $M = 2000$ measurements in a receive interval, and $N_p = 1000$ Nyquist samples and $M_p = 200$ measurements in a pulse width.

For received echo model (4), we assume that the echo amplitudes follow a uniform distribution in $(0,1]$, and the time delays are randomly chosen in resolution grids. Furthermore, we assume that the element of coefficient vector $\boldsymbol{\sigma}$ is nonzero with probability $p$, and zero with probability $(1-p)$. Then the mean sparsity of coefficient vector $\boldsymbol{\sigma}$ is equal to $p$ times the Nyquist samples over the receive time.

### A. *Properties of Virtual Noise*

It is seen that the virtual noise directly affects the reconstruction performance. We now evaluate some properties of the noise.

We first present the existence range of the virtual noise. Note that the length of the virtual measurement in a segment is $\tilde{M} = 800$ when $M_p = 200$ and $S = 3$. Then the forward virtual noise exists in $[1:200]$ and backward virtual noise in $[601:800]$. Fig. 5(a) shows a realization of the virtual noise with respect to the measurement time in the second segment. The result is obtained with $p = 0.005$ and $S = 3$, and (26) is solved by OMP-PKS. It is seen that Fig. 5(a) is consistent with the theoretical analysis.

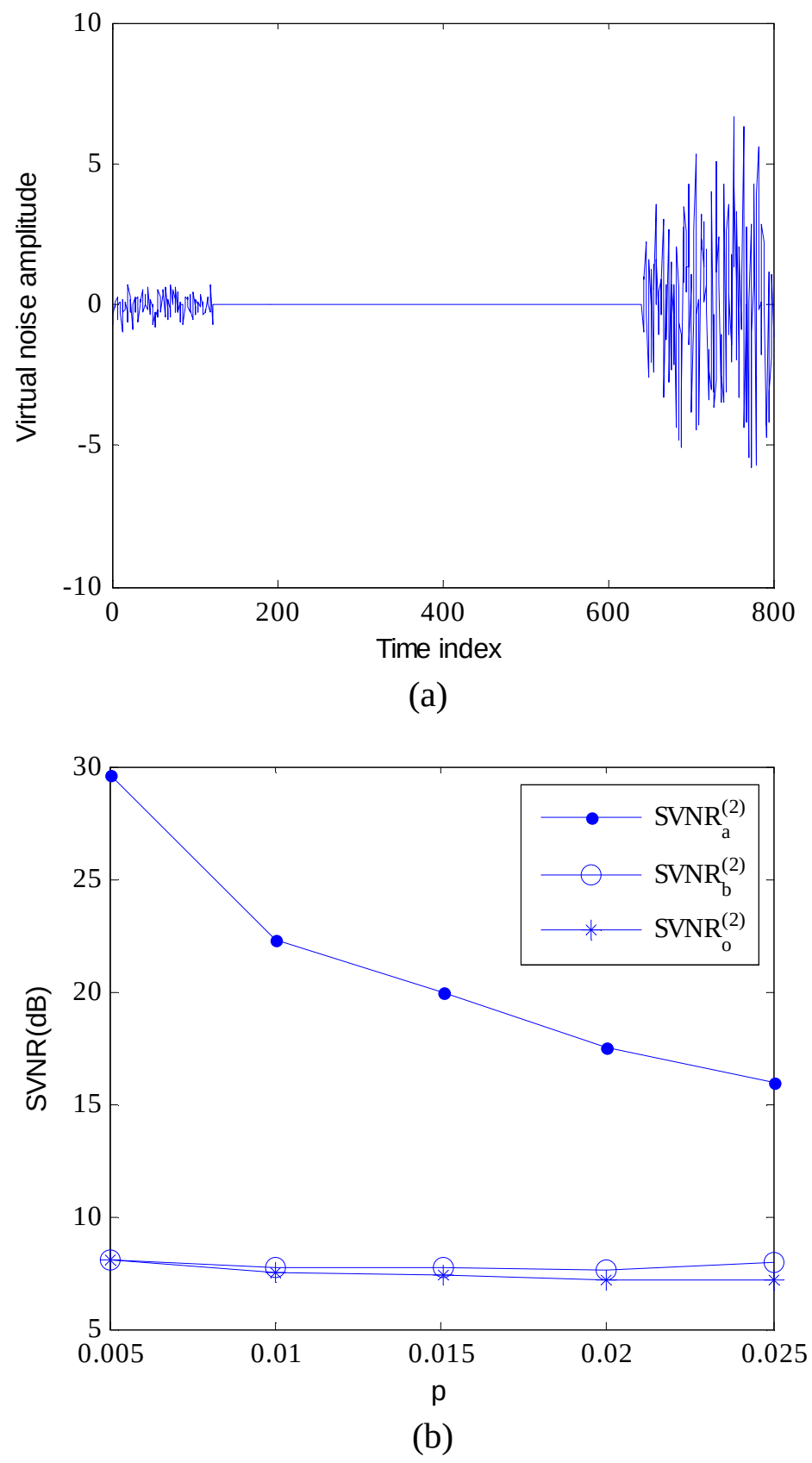

Fig. 5. (a) Variations of virtual noise amplitude. (b) Signal to virtual noise ratio.

Next, we measure the virtual noise level. We define the following three signal-to-noise ratio (SNR) metrics, i.e., $SVNR_o^{(l)}$, $SVNR_a^{(l)}$, and $SVNR_b^{(l)}$, to describe the levels of total virtual noise, forward virtual noise, and backward virtual noise in the $l$-th segment, respectively:

$$SVNR_o^{(l)} = \frac{\left\|\tilde{\mathbf{A}}^{(l)}\tilde{\boldsymbol{\sigma}}^{(l)}\right\|_2^2}{E\left[\left\|\underline{\tilde{\mathbf{n}}}^{(l)}\right\|_2^2\right]}, \quad SVNR_a^{(l)} = \frac{\left\|\tilde{\mathbf{A}}^{(l)}\tilde{\boldsymbol{\sigma}}^{(l)}\right\|_2^2}{E\left[\left\|\underline{\tilde{\mathbf{A}}}^{(l-1)}\Delta\tilde{\boldsymbol{\sigma}}_1^{(l-1)}\right\|_2^2\right]},$$

$$SVNR_b^{(l)} = \frac{\left\|\tilde{\mathbf{A}}^{(l)}\tilde{\boldsymbol{\sigma}}^{(l)}\right\|_2^2}{E\left[\left\|\tilde{\mathbf{A}}^{(l+1)}\tilde{\boldsymbol{\sigma}}_S^{(l+1)}\right\|_2^2\right]}.$$

Fig. 5(b) shows the results of the second segment for different values of $p$, where $S = 3$, and the results are obtained using the OMP-PKS. It is clear that $SVNR_b^{(l)}$ is significantly lower than $SVNR_a^{(l)}$, and $SVNR_o^{(l)}$ is slightly lower than $SVNR_b^{(l)}$. It means that the backward noise level is much larger than that of the forward virtual noise and the virtual noise is dominated by the backward virtual noise.

### B. *Signal Reconstruction in the Noise-Free Case*

We now evaluate the reconstruction error and the running time of our proposed SegSR scheme with OMP-PKS and TOMPP. The effects of different segment lengths and sliding widths are conducted. For simplification, the SegSR scheme with OMP-PKS and TOMPP is simply denoted as OMP-PKS and TOMPP, and the direct reconstruction by OMP is denoted as OMP.

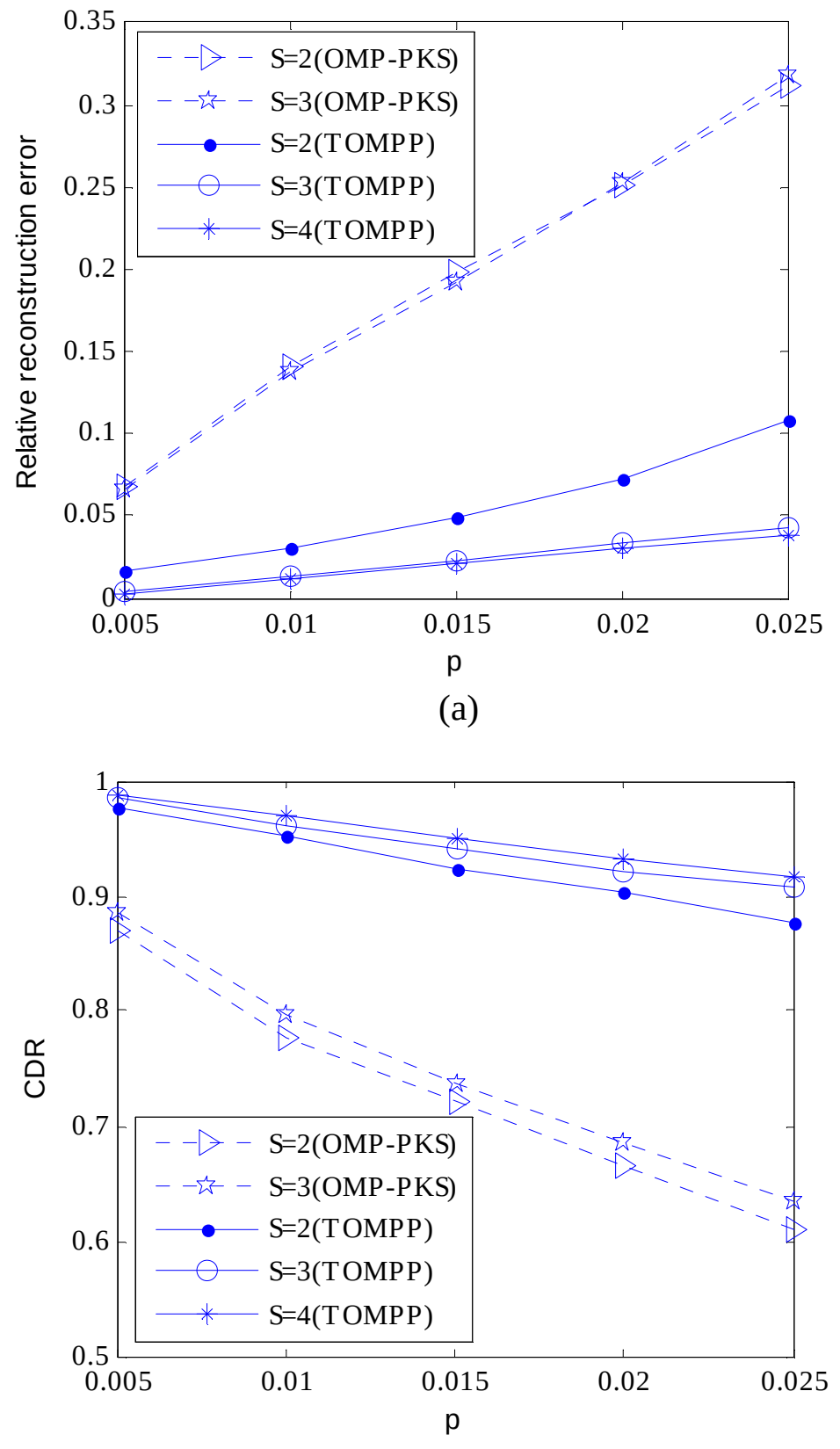

Fig. 6. (a) Relative reconstruction error. (b) Correct discovery rate.

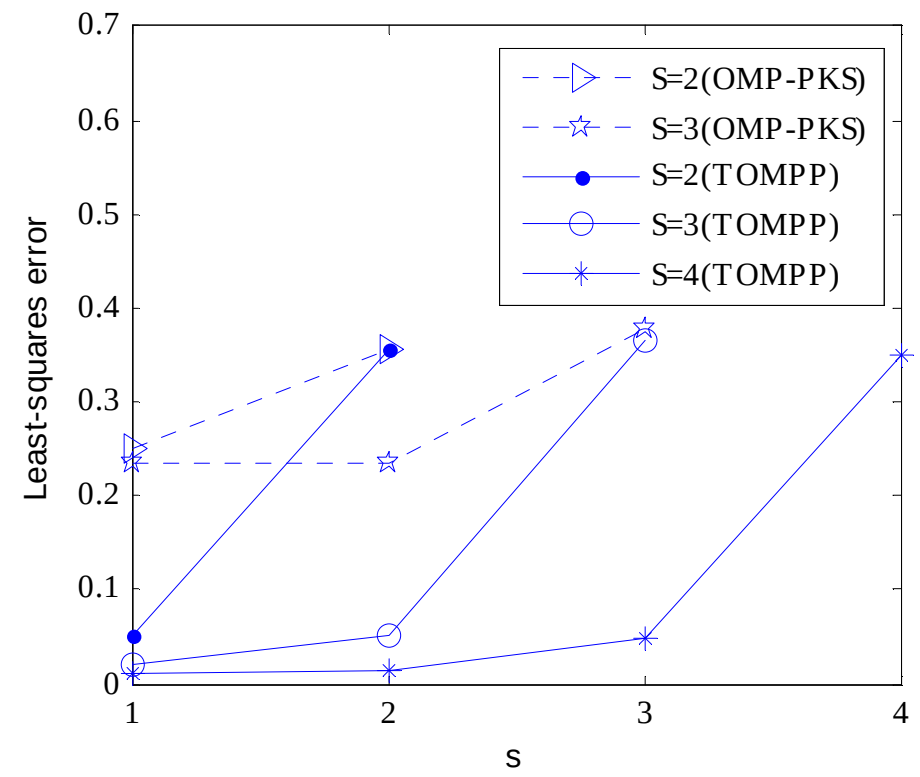

Fig. 7. $\left\|\Delta\tilde{\boldsymbol{\sigma}}_s^{(2)}\right\|_2$ versus $s$.

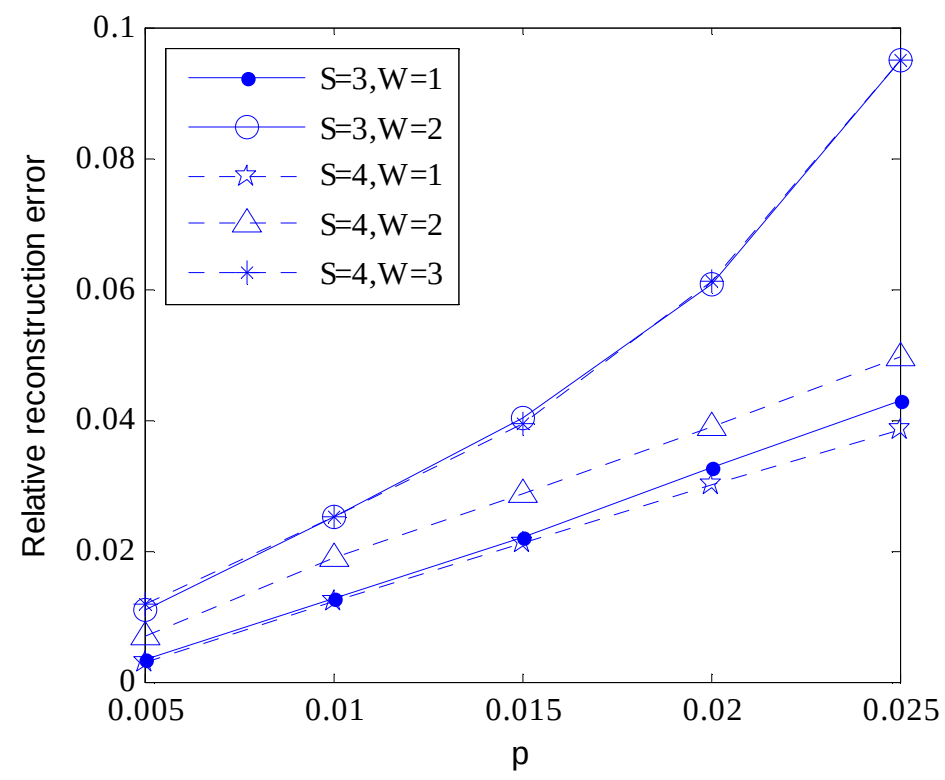

Fig. 8. Relative reconstruction error under different $W$.

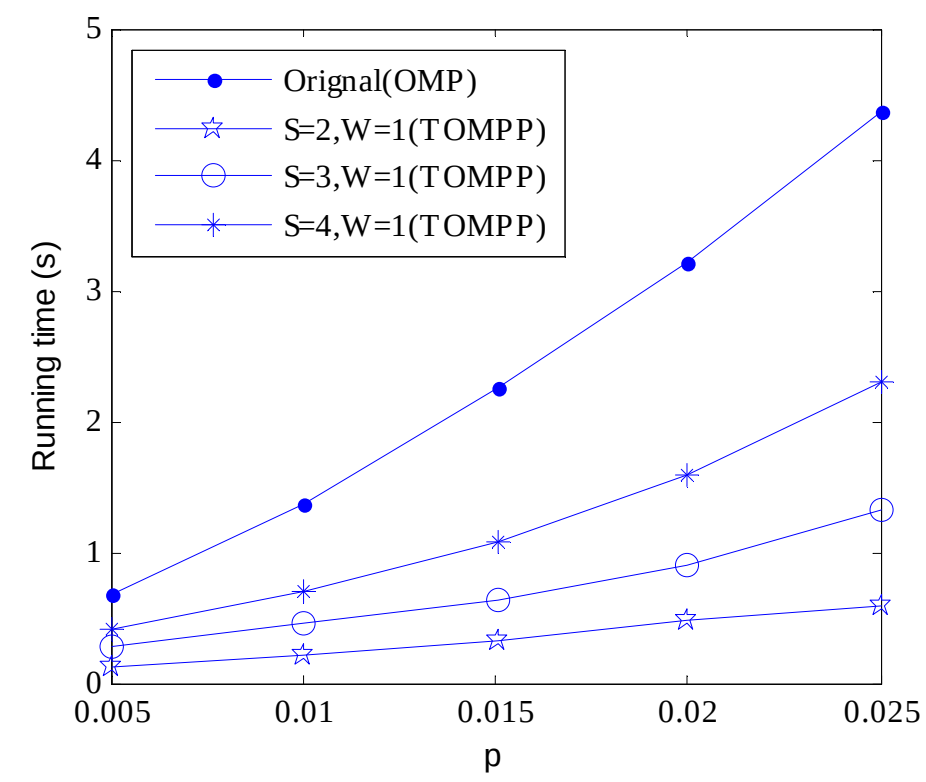

Fig. 9. Running time.

**(1) Reconstruction Error**

We take as performance criterion the relative reconstruction error $E_r$

$$E_r = \frac{\left\|\hat{\boldsymbol{\sigma}} - \boldsymbol{\sigma}\right\|_2}{\left\|\boldsymbol{\sigma}\right\|_2}. \quad (37)$$

Fig. 6(a) shows the relative reconstruction errors versus $p$ for different segment lengths $S$. The reconstruction errors by OMP are not shown here for comparison because they derive much smaller errors in the noise-free case. It is seen that the relative reconstruction errors by TOMPP are much lower than by OMP-PKS. This is because TOMPP can find sparse coefficients with smaller amplitudes than OMP-PKS can do, as we discussed in Section V. Fig. 6(b) further validates the observation through correct discovery rate (CDR) which is defined as the ratio of the number of coefficients correctly declared as nonzero to the total number of nonzero coefficients.

In addition, we notice that the relative reconstruction errors $E_r$ by OMP-PKS remain unchanged for different values of $S$. However, the relative reconstruction error $E_r$ by TOMPP substantially decreases when $S$ is changed from 2 to 3. When $S=4$, the relative reconstruction error $E_r$ does not significantly change from the $S=3$ case. This is because the virtual noise has different effects on the estimate error $\left\|\Delta\tilde{\boldsymbol{\sigma}}_s^{(l)}\right\|_2$ for $s=1,\ldots,S$. To elaborate this, we show in Fig. 7 the variation of $\left\|\Delta\tilde{\boldsymbol{\sigma}}_s^{(l)}\right\|_2$ versus $s$ in the second segment with $p=0.01$. It is seen that $\left\|\Delta\tilde{\boldsymbol{\sigma}}_s^{(l)}\right\|_2$ for $s=1$ is the smallest among all $\left\|\Delta\tilde{\boldsymbol{\sigma}}_s^{(l)}\right\|_2$ ($s=1,\ldots,S$). Meanwhile, $\left\|\Delta\tilde{\boldsymbol{\sigma}}_s^{(l)}\right\|_2$ for $s=1$ and $S=3$ is almost same as for $s=1$ and $S=4$.

Note that the SegSR scheme only outputs the first $N_p$ elements (number of Nyquist samples in a pulse width) in a segment estimation. In practice, we can output the estimates of the first multiple $N_p$ elements, $WN_p$ ($1 \le W < S$). Then, the measurement sub-vector $\tilde{\mathbf{y}}^{(l)}$ should be downward shifted by $WM_p$ compressive samples with respect to $\tilde{\mathbf{y}}^{(l-1)}$. The measurement sub-matrix $\tilde{\mathbf{A}}^{(l)}$ will change correspondingly. Fig. 8 shows the relative reconstruction errors under different $W$ by TOMPP. It is seen that, for a given $p$ and $S$, the relative reconstruction error $E_r$ increases as $W$ increases. That is to say, we obtain the smallest reconstruction error when we downward slide a pulse width. Combined with Fig. 7, TOMPP achieves the optimal performance in terms of reconstruction error when we assign the segment length 4 times pulse width and the sliding length a pulse width.

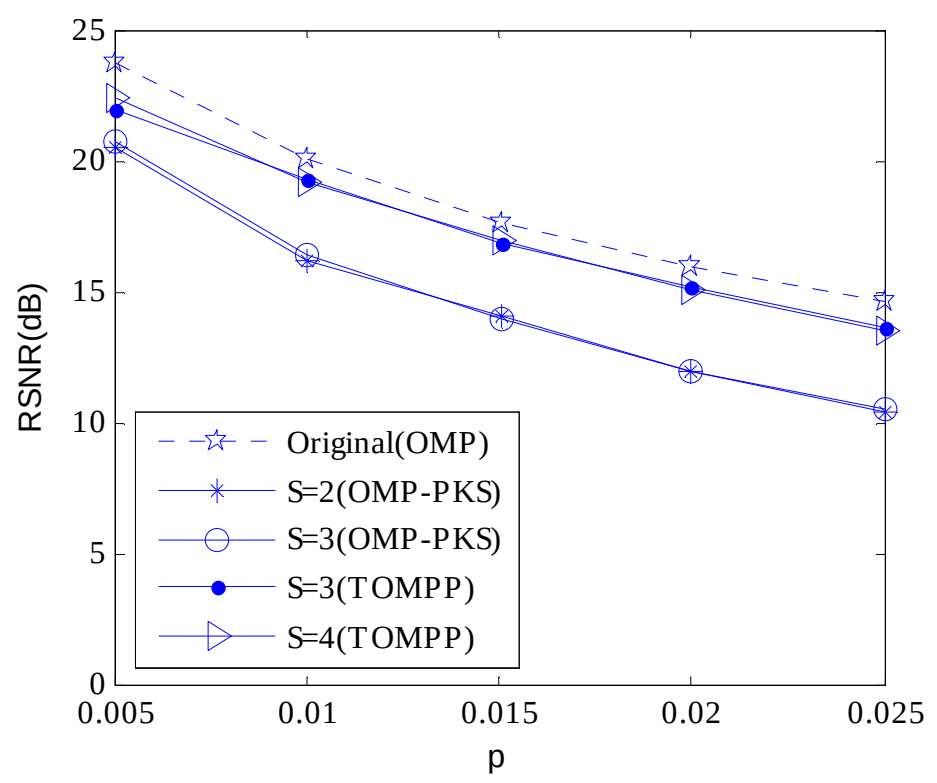

(a)

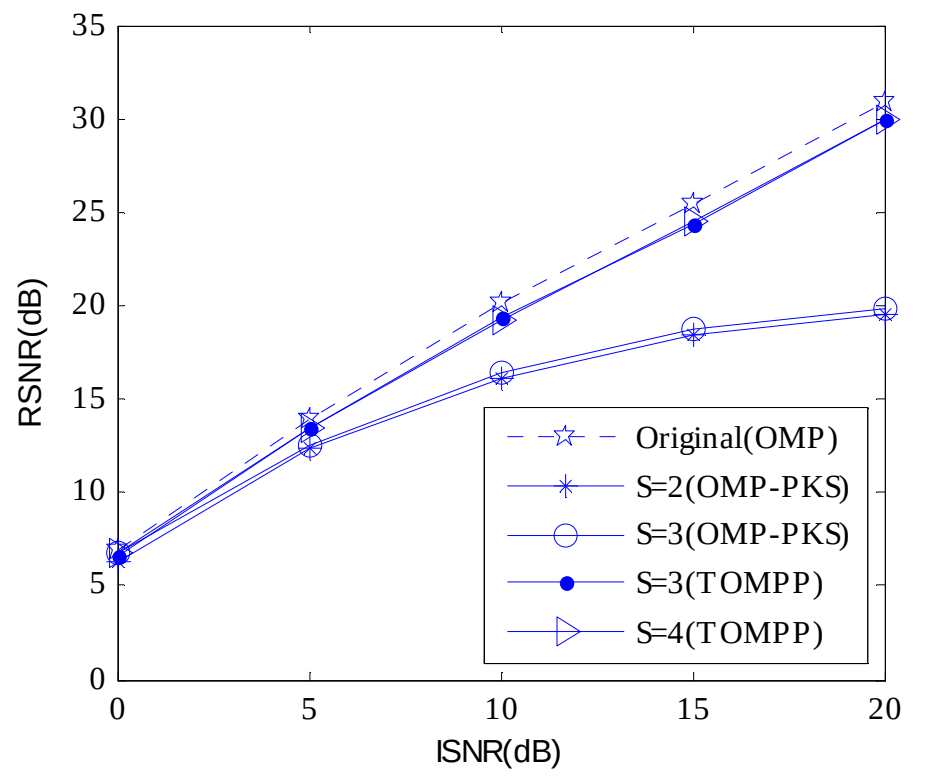

(b)

Fig. 10. (a) RSNR versus $p$ for $\text{ISNR}=10dB$. (b) RSNR versus ISNR for $p=0.01$.

**(2) Running time**

We now use CPU time to illustrate the running time. The simulation is performed in MATLAB 2011b environment on a PC with 3.1GHz Intel core i5-2400 processor and 4 GB RAM. Fig. 9 shows that the TOMPP is faster than the OMP. It should be noted that the simulations are only illustrative for a moderate size example. However, the TOMPP solves large-scale reconstruction problem which cannot be solved by the OMP.

### *C. Signal Reconstruction in the Noise Case*

We assume that the received baseband signal is corrupted by band-limited, additive, white Gaussian noise $n(t)$ with bandwidth $B/2$ and power spectral density $N_0/2$. We use the following reconstruction signal-to-noise ratio (RSNR)

$$RSNR=\frac{\|\boldsymbol{\Psi}\boldsymbol{\sigma}\|_2^2}{E\left[\|\boldsymbol{\Psi}(\hat{\boldsymbol{\sigma}}-\boldsymbol{\sigma})\|_2^2\right]} \tag{38}$$

to evaluate the reconstruction performance.

The variations of RSNR versus $p$ by TOMPP and OMP-PKS are given in Fig. 10(a) for a given input signal-to-noise ratio (ISNR), defined as $ISNR=2\int_0^T|x(t)|^2dt\Big/(TN_0B)$. Fig. 10(b) shows the variations of RSNR versus ISNR by the TOMPP and the OMP-PKS for a given $p$. For comparison, we also give the RSNRs by the OMP. It is seen that the RSNRs by both the TOMPP and the OMP-PKS are smaller than those by the OMP. However, the RSNRs by the TOMPP are close to those by the OMP and are much higher than the OMP-PKS. In case of low ISNRs ($ISNR\le 0\text{dB}$), the OMP, OMP-PKS, and TOMPP all perform poor. This is because they can only find large nonzero coefficients. For high ISNRs, the TOMPP is much better than the OMP-PKS, which further validates the ability of the TOMPP of finding small nonzero coefficients. The simulation results in noisy case further confirm that setting the segment length be 4 is optimal segmenting.

## VIII. Conclusions and Discussions

In this paper, we have developed a SegSR scheme for implementable recovery of full-range pulsed radar echoes from AIC-based sub-Nyquist samples. Our main contributions are summarized as follows:

1) Proper segmentation of the sparse vector and the measurement vector such that the RIP conditions are maintained. This is the core of the proposed SegSR scheme.
2) Analyses of the effect of adjacent segment interference, or virtual noise, on the performance of sparse estimation. Near-optimal length of segments is suggested.
3) A new reconstruction algorithm, TOMPP, for each segment, which considers the virtual noise distributions in the measurement sub-vector and efficiently recovers partial sparse coefficients of each segment.

The proposed SegSR scheme decomposes a large-scale reconstruction problem into a series of small-scale reconstruction ones, thereby significantly reducing the storage and computational load. The SegSR is implemented in a sliding mode and thus is suitable for real-time applications.

This paper takes the RD AIC system as an example to introduce the proposed SegSR scheme. We maintain that it can be applied to other AIC systems in which the measurement matrices have a similar structure as shown in Fig. 1.

## Appendix

### *A. Proof of Theorem 1*

To prove Theorem 1, we introduce some properties of restricted isometry constant of a matrix $\mathbf{A}$, which appear in different publications. The following Lemma 1 gives a summary of these results. One can refer to [3], [19] for more details.

**Lemma 1:** Assume that the matrix $\mathbf{A}$ has restricted isometry constant $\delta_K$. Denote $\Gamma$ and $\Gamma'$ as disjoint support sets of the sparse vectors $\boldsymbol{\sigma}$ and $\boldsymbol{\sigma}'$, respectively, where $|\Gamma|\le k$, $|\Gamma'|\le k'$ and $k+k'\le K$. Then,

(1) $|\langle\mathbf{A}\boldsymbol{\sigma},\mathbf{A}\boldsymbol{\sigma}'\rangle|\le\delta_{k+k'}\|\boldsymbol{\sigma}\|_2\|\boldsymbol{\sigma}'\|_2$,

(2) $\|\mathbf{A}_\Gamma^H\mathbf{A}_{\Gamma'}\|_2\le\delta_K$,

(3) $\left\|\left(\mathbf{A}_\Gamma^H\mathbf{A}_\Gamma\right)^{-1}\right\|_2\le\dfrac{1}{1-\delta_K}$,

(4) for any positive integer $k\le K$, $\delta_k\le\delta_K$.

**Proof of Theorem 1:** First, we show that the virtual noise $\underline{\tilde{\mathbf{n}}}^{(l)}$ is $\ell_2$-bounded. With the definition (24) of $\underline{\tilde{\mathbf{n}}}^{(l)}$, for

$l=2,3,\ldots,L-1$, we have

$$\begin{aligned}\left\|\underline{\tilde{\mathbf{n}}}^{(l)}\right\|_2 &\le \left\|\underline{\tilde{\mathbf{A}}}^{(l-1)}\Delta\tilde{\boldsymbol{\sigma}}_1^{(l-1)}\right\|_2+\left\|\underset{\sim}{\tilde{\mathbf{A}}}^{(l+1)}\tilde{\boldsymbol{\sigma}}_S^{(l+1)}\right\|_2\\ &\le \left\|\tilde{\mathbf{A}}^{(l-1)}\Delta\tilde{\boldsymbol{\sigma}}_1^{(l-1)}\right\|_2+\left\|\tilde{\mathbf{A}}^{(l+1)}\tilde{\boldsymbol{\sigma}}_S^{(l+1)}\right\|_2 \qquad \text{(A.1)}\\ &\le \sqrt{1+\tilde{\delta}^{(l-1)}_{\left|\Delta\Gamma_1^{(l-1)}\right|}}\left\|\Delta\tilde{\boldsymbol{\sigma}}_1^{(l-1)}\right\|_2+\sqrt{1+\tilde{\delta}^{(l+1)}_{\left|\Gamma_S^{(l+1)}\right|}}\left\|\tilde{\boldsymbol{\sigma}}_S^{(l+1)}\right\|_2.\end{aligned}$$

The last inequality is from the RIP condition (2). Define $a_2^{(l-1)}=\sqrt{1+\tilde{\delta}^{(l-1)}_{\left|\Delta\Gamma_1^{(l-1)}\right|}}\left\|\Delta\tilde{\boldsymbol{\sigma}}_1^{(l-1)}\right\|_2$ and $b_2^{(l+1)}=\sqrt{1+\tilde{\delta}^{(l+1)}_{\left|\Gamma_S^{(l+1)}\right|}}\left\|\tilde{\boldsymbol{\sigma}}_S^{(l+1)}\right\|_2$. Then (A.1) can be simply written as

$$\left\|\underline{\tilde{\mathbf{n}}}^{(l)}\right\|_2\le a_2^{(l-1)}+b_2^{(l+1)},\ \ l=2,3,\ldots,L-1. \qquad \text{(A.2)}$$

For $l=1$, $\left\|\underline{\tilde{\mathbf{n}}}^{(1)}\right\|_2=\left\|\underset{\sim}{\tilde{\mathbf{A}}}^{(2)}\tilde{\boldsymbol{\sigma}}_S^{(2)}\right\|_2\le b_2^{(2)}$. For $l=L$, $\left\|\underline{\tilde{\mathbf{n}}}^{(L)}\right\|_2=\left\|\underline{\tilde{\mathbf{A}}}^{(L-1)}\Delta\tilde{\boldsymbol{\sigma}}_1^{(L-1)}\right\|_2\le a_2^{(L-1)}$. Then (27) is proved.

Next, we show that the virtual noise $\underline{\tilde{\mathbf{n}}}^{(l)}$ is $\ell_\infty$-bounded. Similarly, for $l=2,3,\ldots,L-1$, we have

$$\left\|\left(\tilde{\mathbf{A}}^{(l)}\right)^H\underline{\tilde{\mathbf{n}}}^{(l)}\right\|_\infty\le\left\|\left(\tilde{\mathbf{A}}^{(l)}\right)^H\underline{\tilde{\mathbf{A}}}^{(l-1)}\Delta\tilde{\boldsymbol{\sigma}}_1^{(l-1)}\right\|_\infty+\left\|\left(\tilde{\mathbf{A}}^{(l)}\right)^H\underset{\sim}{\tilde{\mathbf{A}}}^{(l+1)}\tilde{\boldsymbol{\sigma}}_S^{(l+1)}\right\|_\infty. \qquad \text{(A.3)}$$

Note that $\underline{\tilde{\mathbf{A}}}^{(l-1)}$ is orthogonal to $\tilde{\mathbf{A}}_s^{(l)}$ $(s=2,3,\ldots,S)$ and $\underset{\sim}{\tilde{\mathbf{A}}}^{(l+1)}$ is orthogonal to $\tilde{\mathbf{A}}_s^{(l)}$ $(s=1,2,\ldots,S-1)$. Then $\left(\tilde{\mathbf{A}}^{(l)}\right)^H\underline{\tilde{\mathbf{A}}}^{(l-1)}=\left(\tilde{\mathbf{A}}_1^{(l)}\right)^H\underline{\tilde{\mathbf{A}}}^{(l-1)}$ and $\left(\tilde{\mathbf{A}}^{(l)}\right)^H\underset{\sim}{\tilde{\mathbf{A}}}^{(l+1)}=\left(\tilde{\mathbf{A}}_S^{(l)}\right)^H\underset{\sim}{\tilde{\mathbf{A}}}^{(l+1)}$. Thus, we have

$$\left\|\left(\tilde{\mathbf{A}}^{(l)}\right)^H\underline{\tilde{\mathbf{n}}}^{(l)}\right\|_\infty\le\left\|\left(\tilde{\mathbf{A}}_1^{(l)}\right)^H\underline{\tilde{\mathbf{A}}}^{(l-1)}\Delta\tilde{\boldsymbol{\sigma}}_1^{(l-1)}\right\|_\infty+\left\|\left(\tilde{\mathbf{A}}_S^{(l)}\right)^H\underset{\sim}{\tilde{\mathbf{A}}}^{(l+1)}\tilde{\boldsymbol{\sigma}}_S^{(l+1)}\right\|_\infty. \qquad \text{(A.4)}$$

We discuss the two terms in the right-hand side of (A.4) separately. For the first term, note that $\left(\tilde{\mathbf{A}}_1^{(l)}\right)^H\underline{\tilde{\mathbf{A}}}^{(l-1)}=\left(\tilde{\mathbf{A}}_2^{(l-1)}\right)^H\tilde{\mathbf{A}}_1^{(l-1)}$. Then, by Property (1) of Lemma 1, we have

$$\begin{aligned}\left\|\left(\tilde{\mathbf{A}}_1^{(l)}\right)^H\underline{\tilde{\mathbf{A}}}^{(l-1)}\Delta\tilde{\boldsymbol{\sigma}}_1^{(l-1)}\right\|_\infty &=\left\|\left(\tilde{\mathbf{A}}_2^{(l-1)}\right)^H\tilde{\mathbf{A}}_1^{(l-1)}\Delta\tilde{\boldsymbol{\sigma}}_1^{(l-1)}\right\|_\infty\\ &=\max_{j\in\{N_p,N_p+1,\ldots,2N_p-1\}}\left|\left\langle\tilde{\mathbf{a}}_j^{(l-1)},\tilde{\mathbf{A}}_1^{(l-1)}\Delta\tilde{\boldsymbol{\sigma}}_1^{(l-1)}\right\rangle\right| \qquad \text{(A.5)}\\ &\le\tilde{\delta}^{(l-1)}_{\left|\Delta\Gamma_1^{(l-1)}\right|+1}\left\|\Delta\tilde{\boldsymbol{\sigma}}_1^{(l-1)}\right\|_2.\end{aligned}$$

Similarly, for the second term, we have

$$\left\|\left(\tilde{\mathbf{A}}_S^{(l)}\right)^H\underset{\sim}{\tilde{\mathbf{A}}}^{(l+1)}\tilde{\boldsymbol{\sigma}}_S^{(l+1)}\right\|_\infty=\left\|\left(\tilde{\mathbf{A}}_{S-1}^{(l+1)}\right)^H\tilde{\mathbf{A}}_S^{(l+1)}\tilde{\boldsymbol{\sigma}}_S^{(l+1)}\right\|_\infty\le\tilde{\delta}^{(l+1)}_{\left|\Gamma_S^{(l+1)}\right|+1}\left\|\tilde{\boldsymbol{\sigma}}_S^{(l+1)}\right\|_2. \qquad \text{(A.6)}$$

Substituting (A.5) and (A.6) into (A.4) results in

$$\left\|\left(\tilde{\mathbf{A}}^{(l)}\right)^H\underline{\tilde{\mathbf{n}}}^{(l)}\right\|_\infty\le\tilde{\delta}^{(l-1)}_{\left|\Delta\Gamma_1^{(l-1)}\right|+1}\left\|\Delta\tilde{\boldsymbol{\sigma}}_1^{(l-1)}\right\|_2+\tilde{\delta}^{(l+1)}_{\left|\Gamma_S^{(l+1)}\right|+1}\left\|\tilde{\boldsymbol{\sigma}}_S^{(l+1)}\right\|_2. \qquad \text{(A.7)}$$

Define $a_\infty^{(l-1)}=\tilde{\delta}^{(l-1)}_{\left|\Gamma_1^{(l-1)}\right|+1}\left\|\Delta\tilde{\boldsymbol{\sigma}}_1^{(l-1)}\right\|_2$ and $b_\infty^{(l+1)}=\tilde{\delta}^{(l+1)}_{\left|\Gamma_S^{(l+1)}\right|+1}\left\|\tilde{\boldsymbol{\sigma}}_S^{(l+1)}\right\|_2$. We have

$$\left\|\left(\tilde{\mathbf{A}}^{(l)}\right)^H\underline{\tilde{\mathbf{n}}}^{(l)}\right\|_\infty\le a_\infty^{(l-1)}+b_\infty^{(l+1)},\ \ l=2,3,\ldots,L-1. \qquad \text{(A.8)}$$

Specifically, for $l=1$, $\left\|\left(\tilde{\mathbf{A}}^{(1)}\right)^H\underline{\tilde{\mathbf{n}}}^{(1)}\right\|_\infty=\left\|\left(\tilde{\mathbf{A}}^{(1)}\right)^H\underset{\sim}{\tilde{\mathbf{A}}}^{(2)}\tilde{\boldsymbol{\sigma}}_S^{(2)}\right\|_\infty\le b_\infty^{(2)}$. For $l=L$, $\left\|\left(\tilde{\mathbf{A}}^{(L)}\right)^H\underline{\tilde{\mathbf{n}}}^{(L)}\right\|_\infty=\left\|\left(\tilde{\mathbf{A}}^{(L)}\right)^H\underline{\tilde{\mathbf{A}}}^{(L-1)}\Delta\tilde{\boldsymbol{\sigma}}_1^{(L-1)}\right\|_\infty\le a_\infty^{(L-1)}$. Then (28) is proved. ■

## B. *Proof of Theorem 4*

Before proceeding, we derive a lemma which is used in the proof of theorem 4.

**Lemma 2:** Let matrix $\mathbf{U}$ be of full column rank and be partitioned as $\mathbf{U}=\left[\mathbf{U}_1,\mathbf{U}_2,\ldots,\mathbf{U}_S\right]$. For $S\ge 3$, assume that $\mathbf{U}_s$ $(s=1,\ldots,S-2)$ is orthogonal to $\mathbf{U}_{s+2}$ $(s=1,\ldots,S-2)$. Then

$$\mathbf{U}^\dagger=\begin{bmatrix}\mathbf{U}_1^\dagger+\ldots+(-1)^{j-1}\mathbf{U}_1^\dagger\prod_{i=2}^{j}\mathbf{U}_i\mathbf{C}_i^\dagger+\ldots+(-1)^{S-1}\mathbf{U}_1^\dagger\prod_{i=2}^{S}\mathbf{U}_i\mathbf{C}_i^\dagger\\ \mathbf{C}_2^\dagger+\ldots+(-1)^{j-1}\mathbf{C}_2^\dagger\prod_{i=3}^{j+1}\mathbf{U}_i\mathbf{C}_i^\dagger+\ldots+(-1)^{S-2}\mathbf{C}_2^\dagger\prod_{i=3}^{S}\mathbf{U}_i\mathbf{C}_i^\dagger\\ \vdots\\ \mathbf{C}_s^\dagger+\ldots+(-1)^{j-1}\mathbf{C}_s^\dagger\prod_{i=s+1}^{j+s-1}\mathbf{U}_i\mathbf{C}_i^\dagger+\ldots+(-1)^{S-s}\mathbf{C}_s^\dagger\prod_{i=s+1}^{S}\mathbf{U}_i\mathbf{C}_i^\dagger\\ \vdots\\ \mathbf{C}_S^\dagger\end{bmatrix}, \qquad \text{(B.1)}$$

where $\mathbf{C}_i=\left(\mathbf{I}-\left[\mathbf{U}_1,\mathbf{U}_2,\ldots,\mathbf{U}_{i-1}\right]\left[\mathbf{U}_1,\mathbf{U}_2,\ldots,\mathbf{U}_{i-1}\right]^\dagger\right)\mathbf{U}_i$, $i=2,\ldots,S$.

***Proof*:** We proof this lemma by mathematical induction.

For $S=2$, it can be shown [35] that

$$\mathbf{U}^\dagger=\left[\mathbf{U}_1,\mathbf{U}_2\right]^\dagger=\begin{bmatrix}\mathbf{U}_1^\dagger-\mathbf{U}_1^\dagger\mathbf{U}_2\mathbf{C}_2^\dagger\\ \mathbf{C}_2^\dagger\end{bmatrix}. \qquad \text{(B.2)}$$

Similarly, for $S=3$, we have

$$\mathbf{U}^\dagger=\left[\mathbf{U}_1,\mathbf{U}_2,\mathbf{U}_3\right]^\dagger=\begin{bmatrix}\left[\mathbf{U}_1,\mathbf{U}_2\right]^\dagger-\left[\mathbf{U}_1,\mathbf{U}_2\right]^\dagger\mathbf{U}_3\mathbf{C}_3^\dagger\\ \mathbf{C}_3^\dagger\end{bmatrix}. \qquad \text{(B.3)}$$

Substituting (B.2) into (B.3) results in

$$\mathbf{U}^\dagger=\begin{bmatrix}\mathbf{U}_1^\dagger-\mathbf{U}_1^\dagger\mathbf{U}_2\mathbf{C}_2^\dagger+\mathbf{U}_1^\dagger\mathbf{U}_2\mathbf{C}_2^\dagger\mathbf{U}_3\mathbf{C}_3^\dagger\\ \mathbf{C}_2^\dagger-\mathbf{C}_2^\dagger\mathbf{U}_3\mathbf{C}_3^\dagger\\ \mathbf{C}_3^\dagger\end{bmatrix}. \qquad \text{(B.4)}$$

Then (B.1) is true when $S=3$.

Assume that (B.1) is true when $S=Q$. Then for $S=Q+1$, by (B.2), we can get

$$\mathbf{U}^\dagger=\begin{bmatrix}\left[\mathbf{U}_1,\mathbf{U}_2,\ldots,\mathbf{U}_Q\right]^\dagger-\left[\mathbf{U}_1,\mathbf{U}_2,\ldots,\mathbf{U}_Q\right]^\dagger\mathbf{U}_{Q+1}\mathbf{C}_{Q+1}^\dagger\\ \mathbf{C}_{Q+1}^\dagger\end{bmatrix}. \qquad \text{(B.5)}$$

Substituting (B.1) with $S=Q$ into (B.5) results in

$$\mathbf{U}^\dagger=\begin{bmatrix}\mathbf{U}_1^\dagger+\ldots+(-1)^{j-1}\mathbf{U}_1^\dagger\prod_{i=2}^{j}\mathbf{U}_i\mathbf{C}_i^\dagger+\ldots+(-1)^{Q}\mathbf{U}_1^\dagger\prod_{i=2}^{Q+1}\mathbf{U}_i\mathbf{C}_i^\dagger\\ \mathbf{C}_2^\dagger+\ldots+(-1)^{j-1}\mathbf{C}_2^\dagger\prod_{i=3}^{j+1}\mathbf{U}_i\mathbf{C}_i^\dagger+\ldots+(-1)^{Q-1}\mathbf{C}_2^\dagger\prod_{i=3}^{Q+1}\mathbf{U}_i\mathbf{C}_i^\dagger\\ \vdots\\ \mathbf{C}_q^\dagger+\ldots+(-1)^{j-1}\mathbf{C}_q^\dagger\prod_{i=q+1}^{j+q-1}\mathbf{U}_i\mathbf{C}_i^\dagger+\ldots+(-1)^{Q-q+1}\mathbf{C}_q^\dagger\prod_{i=q+1}^{Q+1}\mathbf{U}_i\mathbf{C}_i^\dagger\\ \vdots\\ \mathbf{C}_Q^\dagger+(-1)\mathbf{C}_Q^\dagger\mathbf{U}_{Q+1}\mathbf{C}_{Q+1}^\dagger\\ \mathbf{C}_{Q+1}^\dagger\end{bmatrix}.$$

Then (B.1) is true when $S=Q+1$. We thus prove Lemma 2. ■

**Proof of Theorem 4:** We fist prove the bounds (34) for $l=2,3,\ldots,L-1$. With the least-squares estimate (33), the estimate error is given by

$$\begin{aligned}\Delta\tilde{\boldsymbol{\sigma}}^{(l)}_{\Gamma^{(l)}}&=\tilde{\boldsymbol{\sigma}}^{(l)}_{\Gamma^{(l)}}-\hat{\tilde{\boldsymbol{\sigma}}}^{(l)}_{\Gamma^{(l)}}=-\left(\tilde{\mathbf{A}}^{(l)}_{\Gamma^{(l)}}\right)^\dagger\underline{\tilde{\mathbf{n}}}^{(l)}\\ &=-\left(\tilde{\mathbf{A}}^{(l)}_{\Gamma^{(l)}}\right)^\dagger\left(\underline{\tilde{\mathbf{A}}}^{(l-1)}_{\Gamma_1^{(l-1)}}\Delta\tilde{\boldsymbol{\sigma}}^{(l-1)}_{\Gamma_1^{(l-1)}}+\underset{\sim}{\tilde{\mathbf{A}}}^{(l+1)}_{\Gamma_S^{(l+1)}}\tilde{\boldsymbol{\sigma}}^{(l+1)}_{\Gamma_S^{(l+1)}}\right).\end{aligned} \qquad \text{(B.6)}$$

Note that the matrix $\tilde{\mathbf{A}}^{(l)}_{\Gamma^{(l)}}$ is of full column rank and can be partitioned as $\tilde{\mathbf{A}}^{(l)}_{\Gamma^{(l)}}=\left[\tilde{\mathbf{A}}^{(l)}_{\Gamma_1^{(l)}},\tilde{\mathbf{A}}^{(l)}_{\Gamma_2^{(l)}},\ldots,\tilde{\mathbf{A}}^{(l)}_{\Gamma_S^{(l)}}\right]$, and for $S\ge 3$, $\tilde{\mathbf{A}}^{(l)}_{\Gamma_s^{(l)}}$ $(s=1,2,\ldots,S-2)$ is orthogonal to $\tilde{\mathbf{A}}^{(l)}_{\Gamma_{s+2}^{(l)}}$ $(s=1,2,\ldots,S-2)$. Then with Lemma 2, we have

$$\Delta\tilde{\boldsymbol{\sigma}}_{\Gamma_s^{(l)}}^{(l)} = \tilde{\boldsymbol{\sigma}}_{\Gamma_s^{(l)}}^{(l)} - \hat{\tilde{\boldsymbol{\sigma}}}_{\Gamma_s^{(l)}}^{(l)} = -\left(\tilde{\mathbf{P}}_{\Gamma_s^{(l)}}^{(l)} \Delta\tilde{\boldsymbol{\sigma}}_{\Gamma_1^{(l-1)}}^{(l-1)} + \tilde{\mathbf{Q}}_{\Gamma_s^{(l)}}^{(l)} \tilde{\boldsymbol{\sigma}}_{\Gamma_S^{(l+1)}}^{(l+1)}\right), \quad s = 1,\ldots,S, \tag{B.7}$$

where $\tilde{\mathbf{P}}_{\Gamma_s^{(l)}}^{(l)}$ and $\tilde{\mathbf{Q}}_{\Gamma_s^{(l)}}^{(l)}$ are defined at the bottom of the page with $\tilde{\mathbf{C}}_{\Gamma_i^{(l)}}^{(l)} = \left(\mathbf{I} - \left[\tilde{\mathbf{A}}_{\Gamma_1^{(l)}}^{(l)}, \tilde{\mathbf{A}}_{\Gamma_2^{(l)}}^{(l)}, \ldots, \tilde{\mathbf{A}}_{\Gamma_{i-1}^{(l)}}^{(l)}\right]\left[\tilde{\mathbf{A}}_{\Gamma_1^{(l)}}^{(l)}, \tilde{\mathbf{A}}_{\Gamma_2^{(l)}}^{(l)}, \ldots, \tilde{\mathbf{A}}_{\Gamma_{i-1}^{(l)}}^{(l)}\right]^{\dagger}\right)\tilde{\mathbf{A}}_{\Gamma_i^{(l)}}^{(l)}$ for $i = 2,\ldots,S$. Then by (B.7), we have

$$\left\|\Delta\tilde{\boldsymbol{\sigma}}_{\Gamma_s^{(l)}}^{(l)}\right\|_2 \le \left\|\tilde{\mathbf{P}}_{\Gamma_s^{(l)}}^{(l)}\right\|_2 \left\|\Delta\tilde{\boldsymbol{\sigma}}_{\Gamma_1^{(l-1)}}^{(l-1)}\right\|_2 + \left\|\tilde{\mathbf{Q}}_{\Gamma_s^{(l)}}^{(l)}\right\|_2 \left\|\tilde{\boldsymbol{\sigma}}_{\Gamma_S^{(l+1)}}^{(l+1)}\right\|_2, \quad s = 1,\ldots,S. \tag{B.8}$$

We now derive the upper bounds of $\left\|\tilde{\mathbf{P}}_{\Gamma_s^{(l)}}^{(l)}\right\|_2$ and $\left\|\tilde{\mathbf{Q}}_{\Gamma_s^{(l)}}^{(l)}\right\|_2$. From the definitions of $\tilde{\mathbf{P}}_{\Gamma_s^{(l)}}^{(l)}$ and $\tilde{\mathbf{Q}}_{\Gamma_s^{(l)}}^{(l)}$, we can find that $\left\|\tilde{\mathbf{P}}_{\Gamma_s^{(l)}}^{(l)}\right\|_2$ and $\left\|\tilde{\mathbf{Q}}_{\Gamma_s^{(l)}}^{(l)}\right\|_2$ are related with terms $\left\|\left(\tilde{\mathbf{A}}_{\Gamma_1^{(l)}}^{(l)}\right)^{\dagger}\tilde{\mathbf{A}}_{\Gamma_2^{(l)}}^{(l)}\right\|_2$, $\left\|\left(\tilde{\mathbf{A}}_{\Gamma_1^{(l)}}^{(l)}\right)^{\dagger}\underline{\tilde{\mathbf{A}}}_{\Gamma_1^{(l-1)}}^{(l-1)}\right\|_2$, $\left\|\left(\tilde{\mathbf{C}}_{\Gamma_i^{(l)}}^{(l)}\right)^{\dagger}\underline{\tilde{\mathbf{A}}}_{\Gamma_1^{(l-1)}}^{(l-1)}\right\|_2$ $(i = 2,\ldots,S)$, $\left\|\left(\tilde{\mathbf{C}}_{\Gamma_i^{(l)}}^{(l)}\right)^{\dagger}\tilde{\mathbf{A}}_{\Gamma_{i+1}^{(l)}}^{(l)}\right\|_2$ $(i = 2,\ldots,S-1)$ and $\left\|\left(\tilde{\mathbf{C}}_{\Gamma_S^{(l)}}^{(l)}\right)^{\dagger}\underset{\sim}{\tilde{\mathbf{A}}}_{\Gamma_S^{(l+1)}}^{(l+1)}\right\|_2$. Now we discuss these terms separately.

Note that $\left(\tilde{\mathbf{A}}_{\Gamma_1^{(l)}}^{(l)}\right)^{\dagger}\tilde{\mathbf{A}}_{\Gamma_2^{(l)}}^{(l)} = \left(\left(\tilde{\mathbf{A}}_{\Gamma_1^{(l)}}^{(l)}\right)^{H}\tilde{\mathbf{A}}_{\Gamma_1^{(l)}}^{(l)}\right)^{-1}\left(\tilde{\mathbf{A}}_{\Gamma_1^{(l)}}^{(l)}\right)^{H}\tilde{\mathbf{A}}_{\Gamma_2^{(l)}}^{(l)}$. Then for term $\left\|\left(\tilde{\mathbf{A}}_{\Gamma_1^{(l)}}^{(l)}\right)^{\dagger}\tilde{\mathbf{A}}_{\Gamma_2^{(l)}}^{(l)}\right\|_2$, with Properties (2)-(4) of Lemma 1, we get

$$\begin{aligned}\left\|\left(\tilde{\mathbf{A}}_{\Gamma_1^{(l)}}^{(l)}\right)^{\dagger}\tilde{\mathbf{A}}_{\Gamma_2^{(l)}}^{(l)}\right\|_2 &\le \left\|\left(\left(\tilde{\mathbf{A}}_{\Gamma_1^{(l)}}^{(l)}\right)^{H}\tilde{\mathbf{A}}_{\Gamma_1^{(l)}}^{(l)}\right)^{-1}\right\|_2 \left\|\left(\tilde{\mathbf{A}}_{\Gamma_1^{(l)}}^{(l)}\right)^{H}\tilde{\mathbf{A}}_{\Gamma_2^{(l)}}^{(l)}\right\|_2 \\ &\le \frac{1}{1-\tilde{\delta}_{\left|\Gamma^{(l)}\right|}^{(l)}}\tilde{\delta}_{\left|\Gamma^{(l)}\right|}^{(l)} \le \frac{\bar{\delta}_{\bar{K}^{(l)}}^{(l)}}{1-\bar{\delta}_{\bar{K}^{(l)}}^{(l)}}.\end{aligned} \tag{B.9}$$

Similarly, for term $\left\|\left(\tilde{\mathbf{A}}_{\Gamma_1^{(l)}}^{(l)}\right)^{\dagger}\underline{\tilde{\mathbf{A}}}_{\Gamma_1^{(l-1)}}^{(l-1)}\right\|_2$, we have

$$\begin{aligned}\left\|\left(\tilde{\mathbf{A}}_{\Gamma_1^{(l)}}^{(l)}\right)^{\dagger}\underline{\tilde{\mathbf{A}}}_{\Gamma_1^{(l-1)}}^{(l-1)}\right\|_2 &\le \left\|\left(\left(\tilde{\mathbf{A}}_{\Gamma_2^{(l-1)}}^{(l-1)}\right)^{H}\tilde{\mathbf{A}}_{\Gamma_2^{(l-1)}}^{(l-1)}\right)^{-1}\right\|_2 \left\|\left(\tilde{\mathbf{A}}_{\Gamma_2^{(l-1)}}^{(l-1)}\right)^{H}\tilde{\mathbf{A}}_{\Gamma_1^{(l-1)}}^{(l-1)}\right\|_2 \\ &\le \frac{1}{1-\tilde{\delta}_{\left|\Gamma^{(l-1)}\right|}^{(l-1)}}\tilde{\delta}_{\left|\Gamma^{(l-1)}\right|}^{(l-1)} \le \frac{\bar{\delta}_{\bar{K}^{(l)}}^{(l)}}{1-\bar{\delta}_{\bar{K}^{(l)}}^{(l)}}.\end{aligned} \tag{B.10}$$

For term $\left\|\left(\tilde{\mathbf{C}}_{\Gamma_i^{(l)}}^{(l)}\right)^{\dagger}\underline{\tilde{\mathbf{A}}}_{\Gamma_1^{(l-1)}}^{(l-1)}\right\|_2$ $(i = 2,\ldots,S)$, we first derive the bound for $i = 2$. Note that $\underline{\tilde{\mathbf{A}}}^{(l-1)}$ is orthogonal to $\tilde{\mathbf{A}}_s^{(l)}$ $(s = 2,3,\ldots,S)$. Then

$$\left(\tilde{\mathbf{C}}_{\Gamma_2^{(l)}}^{(l)}\right)^{\dagger}\underline{\tilde{\mathbf{A}}}_{\Gamma_1^{(l-1)}}^{(l-1)} = \left(\left(\tilde{\mathbf{C}}_{\Gamma_2^{(l)}}^{(l)}\right)^{H}\tilde{\mathbf{C}}_{\Gamma_2^{(l)}}^{(l)}\right)^{-1}\left(\tilde{\mathbf{A}}_{\Gamma_2^{(l)}}^{(l)}\right)^{H}\tilde{\mathbf{A}}_{\Gamma_1^{(l)}}^{(l)}\left(\tilde{\mathbf{A}}_{\Gamma_1^{(l)}}^{(l)}\right)^{\dagger}\underline{\tilde{\mathbf{A}}}_{\Gamma_1^{(l-1)}}^{(l-1)}. \tag{B.11}$$

Thus

$$\begin{aligned}&\left\|\left(\tilde{\mathbf{C}}_{\Gamma_2^{(l)}}^{(l)}\right)^{\dagger}\underline{\tilde{\mathbf{A}}}_{\Gamma_1^{(l-1)}}^{(l-1)}\right\|_2 \\ &\le \left\|\left(\left(\tilde{\mathbf{C}}_{\Gamma_2^{(l)}}^{(l)}\right)^{H}\tilde{\mathbf{C}}_{\Gamma_2^{(l)}}^{(l)}\right)^{-1}\right\|_2 \left\|\left(\tilde{\mathbf{A}}_{\Gamma_2^{(l)}}^{(l)}\right)^{H}\tilde{\mathbf{A}}_{\Gamma_1^{(l)}}^{(l)}\right\|_2 \left\|\left(\tilde{\mathbf{A}}_{\Gamma_1^{(l)}}^{(l)}\right)^{\dagger}\underline{\tilde{\mathbf{A}}}_{\Gamma_1^{(l-1)}}^{(l-1)}\right\|_2.\end{aligned} \tag{B.12}$$

With the Properties (3) and (4) of Lemma 1, we have

$$\begin{aligned}\left\|\left(\left(\tilde{\mathbf{C}}_{\Gamma_2^{(l)}}^{(l)}\right)^{H}\tilde{\mathbf{C}}_{\Gamma_2^{(l)}}^{(l)}\right)^{-1}\right\|_2 &\le \left\|\left(\left[\tilde{\mathbf{A}}_{\Gamma_1^{(l)}}^{(l)}, \tilde{\mathbf{A}}_{\Gamma_2^{(l)}}^{(l)}\right]^{H}\left[\tilde{\mathbf{A}}_{\Gamma_1^{(l)}}^{(l)}, \tilde{\mathbf{A}}_{\Gamma_2^{(l)}}^{(l)}\right]\right)^{-1}\right\|_2 \\ &\le \frac{1}{1-\tilde{\delta}_{\left|\Gamma^{(l)}\right|}^{(l)}} \le \frac{1}{1-\bar{\delta}_{\bar{K}^{(l)}}^{(l)}}.\end{aligned} \tag{B.13}$$

By Properties (2) and (4) of Lemma 1, we have

$$\left\|\left(\tilde{\mathbf{A}}_{\Gamma_2^{(l)}}^{(l)}\right)^{H}\tilde{\mathbf{A}}_{\Gamma_1^{(l)}}^{(l)}\right\|_2 \le \tilde{\delta}_{\left|\Gamma^{(l)}\right|}^{(l)} \le \bar{\delta}_{\bar{K}^{(l)}}^{(l)}. \tag{B.14}$$

Substituting (B.13), (B.14) and (B.10) into (B.12), we have

$$\left\|\left(\tilde{\mathbf{C}}_{\Gamma_2^{(l)}}^{(l)}\right)^{\dagger}\underline{\tilde{\mathbf{A}}}_{\Gamma_1^{(l-1)}}^{(l-1)}\right\|_2 \le \left(\frac{\bar{\delta}_{\bar{K}^{(l)}}^{(l)}}{1-\bar{\delta}_{\bar{K}^{(l)}}^{(l)}}\right)^2. \tag{B.15}$$

For $i = 3,\ldots,S$,

$$\begin{aligned}&\left\|\left(\tilde{\mathbf{C}}_{\Gamma_i^{(l)}}^{(l)}\right)^{\dagger}\underline{\tilde{\mathbf{A}}}_{\Gamma_1^{(l-1)}}^{(l-1)}\right\|_2 \\ &\le \left\|\left(\left(\tilde{\mathbf{C}}_{\Gamma_i^{(l)}}^{(l)}\right)^{H}\tilde{\mathbf{C}}_{\Gamma_i^{(l)}}^{(l)}\right)^{-1}\right\|_2 \left\|\left(\tilde{\mathbf{A}}_{\Gamma_i^{(l)}}^{(l)}\right)^{H}\tilde{\mathbf{A}}_{\Gamma_{i-1}^{(l)}}^{(l)}\right\|_2 \left\|\left(\tilde{\mathbf{C}}_{\Gamma_{i-1}^{(l)}}^{(l)}\right)^{\dagger}\underline{\tilde{\mathbf{A}}}_{\Gamma_1^{(l-1)}}^{(l-1)}\right\|_2.\end{aligned} \tag{B.16}$$

Similar to (B.13) and (B.14), we have

$$\left\|\left(\left(\tilde{\mathbf{C}}_{\Gamma_i^{(l)}}^{(l)}\right)^{H}\tilde{\mathbf{C}}_{\Gamma_i^{(l)}}^{(l)}\right)^{-1}\right\|_2 \le \frac{1}{1-\tilde{\delta}_{\left|\Gamma^{(l)}\right|}^{(l)}} \le \frac{1}{1-\bar{\delta}_{\bar{K}^{(l)}}^{(l)}}, \tag{B.17}$$

$$\left\|\left(\tilde{\mathbf{A}}_{\Gamma_i^{(l)}}^{(l)}\right)^{H}\tilde{\mathbf{A}}_{\Gamma_{i-1}^{(l)}}^{(l)}\right\|_2 \le \tilde{\delta}_{\left|\Gamma^{(l)}\right|}^{(l)} \le \bar{\delta}_{\bar{K}^{(l)}}^{(l)}. \tag{B.18}$$

Substituting (B.17) and (B.18) into (B.16), we have

$$\begin{aligned}\left\|\left(\tilde{\mathbf{C}}_{\Gamma_i^{(l)}}^{(l)}\right)^{\dagger}\underline{\tilde{\mathbf{A}}}_{\Gamma_1^{(l-1)}}^{(l-1)}\right\|_2 &\le \frac{\bar{\delta}_{\bar{K}^{(l)}}^{(l)}}{1-\bar{\delta}_{\bar{K}^{(l)}}^{(l)}}\left\|\left(\tilde{\mathbf{C}}_{\Gamma_{i-1}^{(l)}}^{(l)}\right)^{\dagger}\underline{\tilde{\mathbf{A}}}_{\Gamma_1^{(l-1)}}^{(l-1)}\right\|_2 \\ &\le \left(\frac{\bar{\delta}_{\bar{K}^{(l)}}^{(l)}}{1-\bar{\delta}_{\bar{K}^{(l)}}^{(l)}}\right)^{i-2}\left\|\left(\tilde{\mathbf{C}}_{\Gamma_2^{(l)}}^{(l)}\right)^{\dagger}\underline{\tilde{\mathbf{A}}}_{\Gamma_1^{(l-1)}}^{(l-1)}\right\|_2, \ i = 3,\ldots,S.\end{aligned} \tag{B.19}$$

---

$$\tilde{\mathbf{P}}_{\Gamma_s^{(l)}}^{(l)} = \begin{cases}\left(\tilde{\mathbf{A}}_{\Gamma_1^{(l)}}^{(l)}\right)^{\dagger}\underline{\tilde{\mathbf{A}}}_{\Gamma_1^{(l-1)}}^{(l-1)} + \sum_{j=2}^{S}\left((-1)^{j-1}\left(\tilde{\mathbf{A}}_{\Gamma_1^{(l)}}^{(l)}\right)^{\dagger}\tilde{\mathbf{A}}_{\Gamma_2^{(l)}}^{(l)}\left(\prod_{i=2}^{j-1}\left(\tilde{\mathbf{C}}_{\Gamma_i^{(l)}}^{(l)}\right)^{\dagger}\tilde{\mathbf{A}}_{\Gamma_{i+1}^{(l)}}^{(l)}\right)\left(\tilde{\mathbf{C}}_{\Gamma_j^{(l)}}^{(l)}\right)^{\dagger}\underline{\tilde{\mathbf{A}}}_{\Gamma_1^{(l-1)}}^{(l-1)}\right), & s = 1, \\ \left(\tilde{\mathbf{C}}_{\Gamma_s^{(l)}}^{(l)}\right)^{\dagger}\underline{\tilde{\mathbf{A}}}_{\Gamma_1^{(l-1)}}^{(l-1)} + \sum_{j=2}^{S-s+1}\left((-1)^{j-1}\left(\prod_{i=s}^{j+s-2}\left(\tilde{\mathbf{C}}_{\Gamma_i^{(l)}}^{(l)}\right)^{\dagger}\tilde{\mathbf{A}}_{\Gamma_{i+1}^{(l)}}^{(l)}\right)\left(\tilde{\mathbf{C}}_{\Gamma_{j+s-1}^{(l)}}^{(l)}\right)^{\dagger}\underline{\tilde{\mathbf{A}}}_{\Gamma_1^{(l-1)}}^{(l-1)}\right), & s = 2,\ldots,S-1, \\ \left(\tilde{\mathbf{C}}_{\Gamma_S^{(l)}}^{(l)}\right)^{\dagger}\underline{\tilde{\mathbf{A}}}_{\Gamma_1^{(l-1)}}^{(l-1)}, & s = S.\end{cases}$$

$$\tilde{\mathbf{Q}}_{\Gamma_s^{(l)}}^{(l)} = \begin{cases}(-1)^{S-1}\left(\tilde{\mathbf{A}}_{\Gamma_1^{(l)}}^{(l)}\right)^{\dagger}\tilde{\mathbf{A}}_{\Gamma_2^{(l)}}^{(l)}\left(\prod_{i=2}^{S-1}\left(\tilde{\mathbf{C}}_{\Gamma_i^{(l)}}^{(l)}\right)^{\dagger}\tilde{\mathbf{A}}_{\Gamma_{i+1}^{(l)}}^{(l)}\right)\left(\tilde{\mathbf{C}}_{\Gamma_S^{(l)}}^{(l)}\right)^{\dagger}\underset{\sim}{\tilde{\mathbf{A}}}_{\Gamma_S^{(l+1)}}^{(l+1)}, & s = 1, \\ (-1)^{S-s}\left(\prod_{i=s}^{S-1}\left(\tilde{\mathbf{C}}_{\Gamma_i^{(l)}}^{(l)}\right)^{\dagger}\tilde{\mathbf{A}}_{\Gamma_{i+1}^{(l)}}^{(l)}\right)\left(\tilde{\mathbf{C}}_{\Gamma_S^{(l)}}^{(l)}\right)^{\dagger}\underset{\sim}{\tilde{\mathbf{A}}}_{\Gamma_S^{(l+1)}}^{(l+1)}, & s = 2,\ldots,S-1, \\ \left(\tilde{\mathbf{C}}_{\Gamma_S^{(l)}}^{(l)}\right)^{\dagger}\underset{\sim}{\tilde{\mathbf{A}}}_{\Gamma_S^{(l+1)}}^{(l+1)}, & s = S.\end{cases}$$

Substituting (B.15) into (B.19), we have

$$\left\| \left( \tilde{\mathbf{C}}^{(l)}_{\Gamma^{(l)}_i} \right)^{\dagger} \tilde{\mathbf{A}}^{(l-1)}_{\Gamma^{(l-1)}_1} \right\|_2 \le \left( \frac{\bar{\delta}^{(l)}_{\bar{K}^{(l)}}}{1-\bar{\delta}^{(l)}_{\bar{K}^{(l)}}} \right)^i, \quad i = 3,\ldots,S. \qquad \text{(B.20)}$$

For term $\left\| \left( \tilde{\mathbf{C}}^{(l)}_{\Gamma^{(l)}_i} \right)^{\dagger} \tilde{\mathbf{A}}^{(l)}_{\Gamma^{(l)}_{i+1}} \right\|_2$ ( $i = 2,\ldots,S-1$ ) and $\left\| \left( \tilde{\mathbf{C}}^{(l)}_{\Gamma^{(l)}_S} \right)^{\dagger} \tilde{\mathbf{A}}^{(l+1)}_{\Gamma^{(l+1)}_S} \right\|_2$, similar to the derivation of (B.20), we have

$$\left\| \left( \tilde{\mathbf{C}}^{(l)}_{\Gamma^{(l)}_i} \right)^{\dagger} \tilde{\mathbf{A}}^{(l)}_{\Gamma^{(l)}_{i+1}} \right\|_2 \le \frac{\bar{\delta}^{(l)}_{\bar{K}^{(l)}}}{1-\bar{\delta}^{(l)}_{\bar{K}^{(l)}}}, \quad i = 2,\ldots,S-1, \qquad \text{(B.21)}$$

$$\left\| \left( \tilde{\mathbf{C}}^{(l)}_{\Gamma^{(l)}_S} \right)^{\dagger} \tilde{\mathbf{A}}^{(l+1)}_{\Gamma^{(l+1)}_S} \right\|_2 \le \frac{\bar{\delta}^{(l)}_{\bar{K}^{(l)}}}{1-\bar{\delta}^{(l)}_{\bar{K}^{(l)}}}. \qquad \text{(B.22)}$$

With (B.9), (B.10) and (B.20)-(B.22), we have $\left\| \tilde{\mathbf{P}}^{(l)}_{\Gamma^{(l)}_s} \right\|_2 \le \alpha^s \left(1-\alpha^{2(S-s+1)}\right) \big/ \left(1-\alpha^2\right)$ and $\left\| \tilde{\mathbf{Q}}^{(l)}_{\Gamma^{(l)}_s} \right\|_2 \le \alpha^{S-s+1}$, $s = 1,\ldots,S$, where $\alpha = \bar{\delta}^{(l)}_{\bar{K}^{(l)}} \big/ \left(1-\bar{\delta}^{(l)}_{\bar{K}^{(l)}}\right)$. Then by (B.8), we have

$$\left\| \Delta\tilde{\boldsymbol{\sigma}}^{(l)}_{\Gamma^{(l)}_s} \right\|_2 \le \beta \left\| \Delta\tilde{\boldsymbol{\sigma}}^{(l-1)}_{\Gamma^{(l-1)}_1} \right\|_2 + \alpha^{S-s+1} \left\| \tilde{\boldsymbol{\sigma}}^{(l+1)}_{\Gamma^{(l+1)}_S} \right\|_2, \quad s = 1,\ldots,S, \qquad \text{(B.23)}$$

where $\beta = \alpha^s \left(1-\alpha^{2(S-s+1)}\right) \big/ \left(1-\alpha^2\right)$, $\alpha = \bar{\delta}^{(l)}_{\bar{K}^{(l)}} \big/ \left(1-\bar{\delta}^{(l)}_{\bar{K}^{(l)}}\right)$.

When $l = 1$, $\Delta\tilde{\boldsymbol{\sigma}}^{(l)}_{\Gamma^{(l)}_s} = -\tilde{\mathbf{Q}}^{(l)}_{\Gamma^{(l)}_s} \tilde{\boldsymbol{\sigma}}^{(l+1)}_{\Gamma^{(l+1)}_S}$ and then $\left\| \Delta\tilde{\boldsymbol{\sigma}}^{(l)}_{\Gamma^{(l)}_s} \right\|_2 \le \alpha^{S-s+1} \left\| \tilde{\boldsymbol{\sigma}}^{(l+1)}_{\Gamma^{(l+1)}_S} \right\|_2$ for $s = 1,\ldots,S$. Similarly, when $l = L$, $\left\| \Delta\tilde{\boldsymbol{\sigma}}^{(l)}_{\Gamma^{(l)}_s} \right\|_2 \le \beta \left\| \Delta\tilde{\boldsymbol{\sigma}}^{(l-1)}_{\Gamma^{(l-1)}_1} \right\|_2$ for $s = 1,\ldots,S$.

This concludes the proof of Theorem 4. ■